\documentclass[twocolumn]{aastex631}

\usepackage{multirow} 
\usepackage{amsmath} 

\received{October 8, 2021}
\revised{October 27, 2021}
\accepted{November 1, 2021}

\shorttitle{$\alpha$-enhancement in simulated SFGs}
\shortauthors{Gebek \& Matthee}
\graphicspath{{./}{Figures/}}

\begin{document}

\title{On the variation in stellar $\alpha$-enhancements of star-forming galaxies in the EAGLE simulation}

\author[0000-0002-0206-8231]{Andrea Gebek}\thanks{E-mail: andrea.gebek@ugent.be}
\affiliation{Department of Physics, ETH Z\"{u}rich \\
Wolfgang-Pauli-Strasse 27, 8093 Z\"{u}rich, Switzerland}
\affiliation{Sterrenkundig Observatorium, Universiteit Gent\\
Krijgslaan 281-S9, B-9000 Gent, Belgium}

\author[0000-0003-2871-127X]{Jorryt Matthee}\thanks{Zwicky Fellow. E-mail: mattheej@phys.ethz.ch}
\affiliation{Department of Physics, ETH Z\"{u}rich \\
Wolfgang-Pauli-Strasse 27, 8093 Z\"{u}rich, Switzerland}



\begin{abstract}

The ratio of $\alpha$-elements to iron in galaxies holds valuable information about the star formation history since their enrichment occurs on different timescales. The fossil record of stars in galaxies has mostly been excavated for passive galaxies, since the light of star-forming galaxies is dominated by young stars which have much weaker atmospheric absorption features. Here we use the largest reference cosmological simulation of the EAGLE project to investigate the origin of variations in stellar $\alpha$-enhancement among star-forming galaxies at $z=0$, and their impact on integrated spectra. The definition of $\alpha$-enhancement  in a composite stellar population is ambiguous. We elucidate two definitions - termed `mean' and `galactic' $\alpha$-enhancement - in more detail. While a star-forming galaxy has a high `mean' $\alpha$-enhancement when its stars formed rapidly, a galaxy with a large `galactic' $\alpha$-enhancement generally had a delayed star formation history. We find that absorption-line strengths of Mg and Fe correlate with variations in $\alpha$-enhancement. These correlations are strongest for the `galactic' $\alpha$-enhancement. However, we show that these are mostly caused by other effects which are cross-correlated with $\alpha$-enhancement, such as variations in the light-weighted age. This severely complicates the retrieval of $\alpha$-enhancements in star-forming galaxies. The ambiguity is not severe for passive galaxies and we confirm that spectral variations in these galaxies are caused by measurable variations in $\alpha$-enhancements. We suggest that this more complex coupling between $\alpha$-enhancement and star formation histories can guide the interpretation of new observations of star-forming galaxies.

\end{abstract}



\section{Introduction}

Stars enrich the ambient interstellar medium (ISM) with metals through stellar winds and supernovae (SN) explosions. Elements formed in the $\alpha$-process (the $\alpha$-elements C, O, Ne, Mg, Si, S, Ar, and Ca) are predominantly produced by Type II SNe, which mark the endpoint of the stellar evolution of massive, short-lived stars. On the other hand, the enrichment in iron-peak elements (Fe, Cr, Co, Ni, Cu, Mn) receives a significant contribution from Type Ia SNe (e.g. \citealt{Tinsley1979}; \citealt{GreggioRenzini1983}; \citealt{Wiersma2009b}). These occur in binaries when a white dwarf accretes a critical amount of material from its companion star. Since these SNe explosions exhibit significantly different timescales, the ratio of $\alpha$-elements to iron-peak elements contains information about the star formation history (SFH) of a composite stellar population such as a galaxy. This ratio is often parametrised by the $\alpha$-enhancement [$\alpha$/Fe]\footnote{We adopt the following convention for stellar abundance ratios: [A/B]$=\log_{10}(X^{\mathrm{A}}/X^{\mathrm{B}})-\log_{10}(X^{\mathrm{A}}_{\odot}/X^{\mathrm{B}}_{\odot})$, where $X^j$ denotes the abundance (stellar mass fraction) of element $j$.}, where $\alpha$ comprises all $\alpha$-elements.

It is well known that the delayed iron enrichment from SNe Ia leads to a declining stellar $\alpha$-enhancement\footnote{We only analyse stellar $\alpha$-enhancement throughout this work, note that we usually omit `stellar' in the subsequent text.} over cosmic time \citep[e.g.][]{Weinberg2017}. At the same time, the stellar metallicities of galaxies generally increase as each generation of stars returns a fraction of their mass as metals into gas in the ISM, which then partly fuels later generations of stars. Therefore, the distribution of stars in the [$\alpha$/Fe]-[Fe/H] plane may act as a chemical fossil record of the star formation and chemical enrichment history of a galaxy \citep[e.g.][]{Mackereth2018}.

 For simple stellar populations (SSPs) such as globular clusters, there is a wealth of evidence both from observations and simulations that the timescale of star formation and $\alpha$-enhancement are indeed causally related \citep{Puzia2005,Woodley2010,Hughes2020}. In our own Milky Way, recent large spectroscopic campaigns (e.g. Gaia-ESO, \citealt{Gilmore2012}; GALAH, \citealt{DeSilva2015} and \citealt{Martell2017}; APOGEE, \citealt{Majewski2017}) have made it possible to map the distribution of individual stars in the [$\alpha$/Fe]-[Fe/H] plane \citep[e.g.][]{Hayden2015}. This plane is bimodal and contains a high- and a low-$\alpha$-sequence (e.g. \citealt{Ratcliffe2020}; \citealt{Vincenzo2021}). These sequences are often associated with distinct structures which formed at different times sharing orbital parameters and are therefore used to obtain a detailed picture of the history of our Galaxy \citep[e.g.][]{FreemanBlandHawthorn2002,Venn2004,Naidu2020,Spitoni2021}.

Extending such studies to distant galaxies is challenging. Measurements of $\alpha$-enhancement in integrated spectra have so far mainly been limited to passive galaxies out to a redshift of $z\approx1$ \citep[e.g.][]{Thomas2005,Thomas2010,Rosa2011,Conroy2014,McDermid2015,Kriek2019}. These studies either base their measurements of $\alpha$-enhancement on a combination of Lick indices, or in the case of \cite{Conroy2014} full spectral fitting. In a pioneering study of passive galaxies, \citet{Thomas2005} used a chemical evolution model to relate $\alpha$-enhancement to the full width at half maximum (FWHM) $\Delta t$ of a gaussian-shaped star formation history following $[\alpha/\mathrm{Fe}]\approx1/5-1/6\log_{10}(\Delta t)$. Subsequently, \citet{Rosa2011} found an empirical relation between the timescale of the SFH and galactic $\alpha$-enhancement, as predicted by chemical evolution models. Multiple studies measured the $\alpha$-enhancement of samples of passive galaxies, where a notable trend with galactic stellar mass was found: $\alpha$-enhanced galaxies tend to be more massive, consistent with the idea of `galactic downsizing' (massive galaxies tend to form their stars earlier and more rapidly). This trend, starting at $M_*\approx10^{10.5}M_{\odot}$, has been reproduced in the EAGLE simulation (\citealt{Segers2016}). These authors attribute the rapid quenching of massive galaxies to the feedback of active galactic nuclei (AGN). 

It is of interest to use measurements of $\alpha$-enhancements to constrain the star formation histories of star-forming galaxies (SFGs) on relatively long timescales. Such measurements would serve as interesting cross-checks for recent model-based  \citep[e.g.][]{Pacifici2013,Pacifici2016} or non-parametric \citep[e.g.][]{Iyer2019,Leja2019} inferences of SFHs which are crucial for obtaining reliable stellar masses \citep[e.g.][]{Leja2019b}. They may also help extending the dynamic range of observations of the power spectrum of the time-variability of star formation histories \citep[e.g.][]{Tacchella2020}.  However, measurements of $\alpha$-enhancement for star-forming galaxies are much more challenging than those for passive galaxies as the atmospheres of the young stars that dominate the spectra of such galaxies have much weaker absorption features compared to the stars in passive galaxies \citep{Walcher2009,Conroy2013}. \cite{Gallazzi2021} however recently presented first measurements of $\alpha$-enhancement in the star-forming galaxy population in the local Universe. Furthermore, resolved observations of local star-forming galaxies are also starting to report measurements of $\alpha$-enhancement in star-forming galaxies \citep[e.g.][]{Pinna2019,Neumann2020,Sanchez2020}. A detailed investigation of the relation between stellar $\alpha$-enhancement and the star formation histories of star-forming galaxies is therefore timely.

Besides the observational challenges, the interpretation of observed $\alpha$-enhancements in star-forming galaxies is not trivial. Most observational studies of $\alpha$-enhancement in integrated galaxy spectra treat the galaxies as simple stellar populations. For passive galaxies, this is justifiable by the fact that the (light-weighted) spectrum of an old stellar population that formed in a narrow and rapid star formation history resembles a stellar population with a single age. As we show in this work, this simplification can not be applied to star-forming galaxies which tend to have a longer and slower star formation history.

In this paper, we use the largest reference run of the suite of cosmological hydrodynamical simulations of the EAGLE project (henceforth referred to as `EAGLE simulation') to investigate the origin of stellar $\alpha$-enhancement in terms of galactic star formation histories and chemical evolution. We aim to identify the properties that drive galactic $\alpha$-enhancement in the context of star-forming galaxies, and restrict ourselves to the stellar $\alpha$-enhancement (instead of the gas-phase $\alpha$-enhancement). We use the same set of solar abundances as in \citet{Wiersma2009b}, with the following mass fractions: $X_{\odot}^{\rm{H}}=0.7065$, $X_{\odot}^{\rm{M}}=0.0127$, $X_{\odot}^{\rm{O}}=5.49\times10^{-3}$, $X_{\odot}^{\rm{Fe}}=1.10\times10^{-3}$, $X_{\odot}^{\rm{Mg}}=5.91\times10^{-4}$, with $X_{\odot}^{\rm{M}}$ denoting the total metal mass fraction of the Sun.

We present two distinct definitions of the average galactic $\alpha$-enhancement in \S~\ref{sec:Caution}, and summarize the relevant aspects of the EAGLE simulation and our galaxy sample in \S~\ref{sec:Simulations}. We relate these $\alpha$-enhancements to galactic star formation histories in \S~\ref{sec:Evolution} and discuss how variations in the $\alpha$-enhancements affect the integrated spectral energy distributions in \S~\ref{sec:Impact}. We investigate $\alpha$-enhancement in passive galaxies and compare these results to the ones obtained for star-forming galaxies in \S~\ref{sec:Discussion}, discuss future improvements in \S~\ref{sec:Future} and summarise our work in \S~\ref{sec:Conclusions}.

\section{Defining $\alpha$-enhancement: a cautionary note}\label{sec:Caution}

Given that one knows the chemical composition of a star or SSP, the calculation of [$\alpha$/Fe] is trivial as it is a simple abundance ratio. However, for a composite stellar population (CSP) whose members have individually varying [$\alpha$/Fe], multiple ways of defining and calculating the average CSP $\alpha$-enhancement exist. We illustrate this `averaging ambiguity' with a hypothetical quantity $x$, which is defined as the ratio of $a$ and $b$: $x=a/b$. Assume that we have a population whose members feature varying $a$ and $b$, and we want to calculation a population-average $x$ value. Mathematically, two ways to compute this average exist: $\langle a\rangle/\langle b\rangle$ and $\langle a/b\rangle$, where the brackets denote averaging over the population. These two quantities differ depending on the distributions of $a$ and $b$ in the population, and it is not a priori clear which definition to pick.

This averaging ambiguity naturally arises when averaging (which is a linear operation) a quantity which is a non-linear composition of features, as is the $\alpha$-enhancement [$\alpha$/Fe]. Each different operational ordering of averaging and non-linear operations leads to a distinct population-averaged quantity, such that the number of possible definitions is one plus the number of non-linear operations. Since the $\alpha$-enhancement exhibits two non-linear operations in its definition (ratio and logarithm), three \textit{mathematically equivalent} population-averaged definitions of [$\alpha$/Fe] exist.

\subsection{Average $\alpha$-enhancement of a Composite Stellar Population}\label{sec:AverageAlpha}

From the mathematical viewpoint of how the operations are ordered when calculating [$\alpha$/Fe] for a CSP, three distinct definitions exist. This is reminiscent of the different ages (mass- and light-weighted) that can be assigned to a CSP, however, the various weighting schemes are completely independent of the operational ordering of the averaging process. Hence, we end up with six physically plausible [$\alpha$/Fe] definitions for a CSP, as we have the mass- and light-weighted versions of three operational orderings. The critical point for the astrophysical relevance of these various definitions is their measurability from galactic integrated light spectra. Put another way, what do we actually measure when retrieving [$\alpha$/Fe] from a galactic spectrum?

Pioneering studies measured [$\alpha$/Fe] in early-type galaxies by fitting the observed magnesium and iron Lick indices to SSP models (e.g. \citealt{Trager2000}; \citealt{Thomas2005}; \citealt{Thomas2010}; \citealt{Rosa2011}; \citealt{Johansson2012}; \citealt{McDermid2015}). \citet{TragerSomerville2009} show that such SSP-equivalent retrievals can recover mass- and light-weighted metallicities, but are biased for ages. In the case of $\alpha$-enhancement of a galaxy, we naively expect to retrieve the following quantity:

\begin{equation}\label{eq:obs1}
\begin{split}
    [\alpha/\mathrm{Fe}]_{\mathrm{obs1}}&=\log_{10}\Bigg(\frac{\big\langle X^{\alpha} \big\rangle_L}{\big\langle X^{\mathrm{Fe}} \big\rangle_L}\Bigg)-\log_{10}\Biggl(\frac{X^{\alpha}_{\odot}}{X^{\mathrm{Fe}}_{\odot}}\Biggr)\\
    &=\log_{10}\Biggl(\frac{\sum_i L_i\times X^{\alpha}_i}{\sum_i L_i\times X^{\mathrm{Fe}}_i}\Biggr)-\log_{10}\Biggl(\frac{X^{\alpha}_{\odot}}{X^{\mathrm{Fe}}_{\odot}}\Biggr),
\end{split}
\end{equation}

where $X_i^{j}$ ($X_{\odot}^{j}$) denotes the abundance or mass fraction (we use these two terms interchangeably) of element $j$ in an SSP $i$ (in the Sun). We abbreviate averages with their weights $y$ by $\langle x\rangle_y$, $L_i$ denotes the luminosity of an individual SSP.

More recent observational studies of extragalactic [$\alpha$/Fe] (e.g. \citealt{Pinna2019}; \citealt{Neumann2020}) often employ spectral energy distribution (SED) fitting codes (e.g. pPXF \citealt{Cappellari2004}, Steckmap \citealt{Ocvirk2006}, Starlight \citealt{CidFernandes2005}) to decompose the stellar component of a galactic spectrum into a linear combination of SSPs: $F_{\nu,\mathrm{int}}=\sum_i L_i\times F_{\nu}(t_i,Z_i,[\alpha/\mathrm{Fe}]_i)$. The output of these SED fitting codes consists of the SSP parameters ($t_i,Z_i,[\alpha/\mathrm{Fe}]_i)$) and their luminosity weights ($L_i$) which enter the linear combination for the galactic SED ($F_{\nu,\mathrm{int}}$). This approach uses more information encoded in the galactic SED compared to analyses based on spectral indices and enables the recovery of non-parametric star formation histories.  Note that since the mass-to-light ratios of the SSP models are usually known it is trivial to convert the luminosity into mass weights, such that the retrieved average galactic $\alpha$-enhancement takes the following form:

\begin{equation}\label{eq:obs2}
\begin{split}
    [\alpha/\mathrm{Fe}]_{\mathrm{obs2}}=\langle [\alpha/\mathrm{Fe}]\rangle_m & =\Bigg\langle\log_{10}\Big(\frac{X^{\alpha}}{X^{\mathrm{Fe}}}\Big)\Bigg\rangle_m \\& -\log_{10}\Biggl(\frac{X^{\alpha}_{\odot}}{X^{\mathrm{Fe}}_{\odot}}\Biggr),
\end{split}
\end{equation}

where the SSP $\alpha$-enhancements are weighted by their stellar mass $m$.

Comparing Eqns.~\ref{eq:obs1} and~\ref{eq:obs2}, we find that a galaxy can have different average $\alpha$-enhancements depending on the retrieval method. In a theoretical study of $\alpha$-enhancement where the stellar populations of a galaxy are known a priori (as is the case in this work), one can calculate all six possible average galactic [$\alpha$/Fe]. Here we analyse two specific definitions for the following reasons: we mostly consider mass-weighted $\alpha$-enhancements as those are more interesting for chemical evolution studies of galaxies. Hence, we define the `galactic $\alpha$-enhancement' which corresponds to Definition~\ref{eq:obs1} but using mass weights:

\begin{equation}\label{eq:alpha_gal}
    [\alpha/\mathrm{Fe}]_{\mathrm{gal}}=\log_{10}\Bigg(\frac{\big\langle X^{\alpha} \big\rangle_m}{\big\langle X^{\mathrm{Fe}} \big\rangle_m}\Bigg)-\log_{10}\Biggl(\frac{X^{\alpha}_{\odot}}{X^{\mathrm{Fe}}_{\odot}}\Biggr).
\end{equation}
$[\alpha/\mathrm{Fe}]_{\mathrm{gal}}$ corresponds to the ratio of the galactic $\alpha$-element to the galactic iron abundance. The second [$\alpha$/Fe] definition which we analyse in this study is the `mean $\alpha$-enhancement', which slightly differs from Definition~\ref{eq:obs2}:

\begin{equation}\label{eq:alpha_mean}
    [\alpha/\rm{Fe}]_{\mathrm{mean}}=\log_{10}\Bigg(\bigg\langle \frac{X^{\alpha}}{ X^{\mathrm{Fe}}}\bigg\rangle_m\Bigg)-\log_{10}\Biggl(\frac{X^{\alpha}_{\odot}}{X^{\mathrm{Fe}}_{\odot}}\Biggr).
\end{equation}
$[\alpha/\mathrm{Fe}]_{\mathrm{mean}}$ corresponds to the mean mass-weighted $\alpha$-enhancement of the stars within in a galaxy.
[$\alpha/\mathrm{Fe}]_{\mathrm{mean}}$ and [$\alpha/\mathrm{Fe}]_{\mathrm{obs2}}$ behave very similarly, but [$\alpha/\mathrm{Fe}]_{\mathrm{mean}}$ is more readily comparable to [$\alpha/\mathrm{Fe}]_{\mathrm{gal}}$ which eases the joint analysis of the two quantities.

\begin{figure*}
	\includegraphics[width=\textwidth]{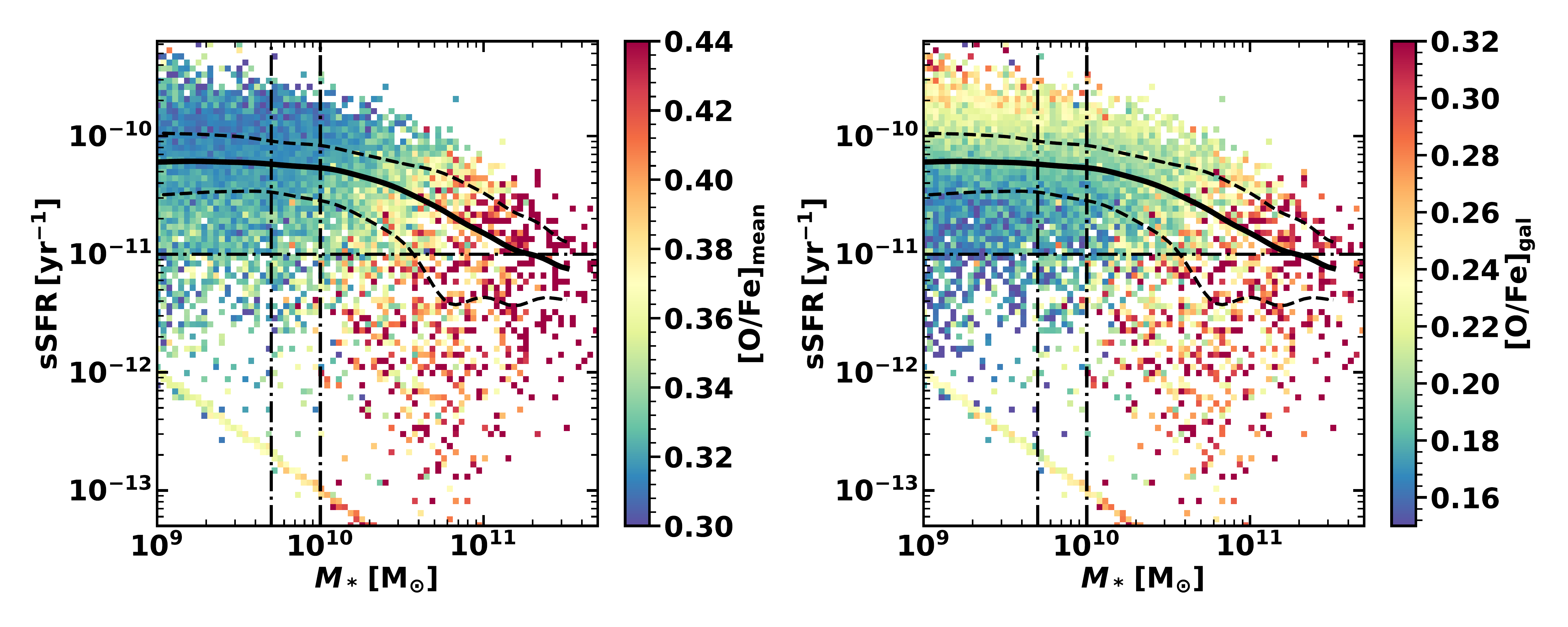}
    \caption{Relation between specific star formation rate and stellar mass for galaxies in the EAGLE simulation at $z=0$. Galaxies are color-coded by two different definitions of $\alpha$-enhancement, by $\mathrm{[O/Fe]_{mean}}$ (Eqn.~\ref{eq:alpha_mean}) in the left panel and by $\mathrm{[O/Fe]_{gal}}$ (Eqn.~\ref{eq:alpha_gal}) in the right panel. The $\alpha$-enhancements are calculated on a logarithmic $100\times100$-grid, the color represents the average (weighted by $M_*$) mean or galactic $\alpha$-enhancements for each cell. The range of the colorbar is limited to better visualise the trends. The galaxies on the diagonal sequence in the bottom-left are devoid of star forming gas and formally have a zero star formation rate. We display them with a constant $\rm{SFR}=10^{-3}$ M$_{\odot}$ yr$^{-1}$ for visualisation  purposes. Dash-dotted black lines show the mass and (sSFR) cuts we use to select the galaxy sample (\S~\ref{sec:GalaxySample}). The solid black lines indicate the running median sSFR of star-forming galaxies (the galactic main sequence), the dashed lines show the interquartile range.}
    \label{fig:MainSequence}
\end{figure*}

\section{Simulation Data}\label{sec:Simulations}

We study galaxies that have been simulated in the cosmological hydrodynamical EAGLE simulation suite (\citealt{Schaye2015}; \citealt{Crain2015}). Here we briefly describe the main code and its subgrid physics, with emphasis on the chemical enrichment scheme (\S~\ref{sec:ChemicalEnrichmentScheme}). We also specify the sample that we selected in our study and how this compares to the total sample of galaxies at $z=0$.

\subsection{Simulation methods}
The EAGLE simulation used throughout this work is based on a standard $\Lambda$CDM-Universe with the cosmological parameters from the 2013 \textit{Planck} mission (\citealt{Planck2014}). Furthermore, we use 13.798\,Gyr for the age of the Universe based on the 2015 \textit{Planck} results (\citealt{Planck2016}). EAGLE is run using a heavily modified version of the $N$-body smoothed particle hydrodynamics (SPH) code \textsc{gadget3}, last described by \citet{Springel2005}. For this work we use the largest reference simulation Ref-L0100N1504 to which we refer as `EAGLE simulation'. This simulation has a box size of $100\,\mathrm{cMpc}$ and contains $2\times1504^3$ baryonic \& dark matter particles. The resolution of the baryonic particles is chosen to marginally resolve the Jeans scale in the warm ISM (\citealt{Schaye2015}). Haloes are identified using a friends-of-friends algorithm (\citealt{Davis1985}) for the dark matter particles. Subsequently, the \textsc{subfind} algorithm (\citealt{Springel2001}; \citealt{Dolag2009}) finds substructures bound by saddle points in the density distribution. These gravitationally bound subhaloes are then identified as galaxies. Properties of the baryonic \& dark matter particles (\citealt{EAGLE2017}), and of the galaxies (\citealt{McAlpine2016}), are stored in 29 snapshots between redshift 20 and 0.

The baryonic particles, which occur in different phases (gas, stars, black holes) have masses of the order $\sim\!10^6\,M_{\odot}$ for the reference model we are using. Relevant processes which are not resolved by the simulation are implemented as subgrid physics. Radiative cooling is calculated using the 11 most important elements for the cooling function: H, He, C, N, O, Ne, Mg, Si, S, Ca, and Fe (\citealt{Wiersma2009a}). Hydrogen is reionized instantaneously at $z=11.5$ when the ionizing background of \citet{Haardt2001} is turned on. Star formation is implemented according to \citet{Schaye2008} with a star formation rate (SFR) following the empirical Kennicutt-Schmidt relation (\citealt{Kennicutt1998}). The masses of newly formed simple stellar populations (SSPs) are distributed according to a Chabrier initial mass function for individual stars with a logarithmic slope of 1.3 for the high-mass end (\citealt{Chabrier2003}; Eqn. 8 in \citealt{Maschberger2013}) within the mass range $0.1-100\,M_{\odot}$. Star formation feeds energy back into the ISM, through both stellar winds and core-collapse supernovae. This feedback is thermal and implemented stochastically to minimise numerical losses (\citealt{DallaVecchia2008}). In EAGLE, feedback from supermassive black holes contributes to the quenching of star formation in massive galaxies (\citealt{Booth2009}). The parameters associated with these feedback processes are chosen such that the simulation simultaneously reproduces $z\sim0$ observations of the galaxy stellar mass function, galaxy sizes, and the relation between black hole mass and stellar mass (\citealt{Schaye2015}; \citealt{Crain2015}). The EAGLE simulation is able to reproduce many observables beyond the aforementioned calibration relations, including the evolution of the galaxy stellar mass function \citep{Furlong2015}, the stellar mass-specific star formation rate (sSFR) sequence, the Tully-Fisher relation (\citealt{Tully1977}), and galactic mass-metallicity relations \citep{DeRossi2017}.

\subsection{Chemical Enrichment Scheme}\label{sec:ChemicalEnrichmentScheme}

Since the $\alpha$-enhancement is determined by the chemical enrichment prescription within EAGLE, we describe the simulated processes here. A general enrichment scheme within SPH-codes is described in \citet{Wiersma2009b}, which is also used for the EAGLE simulation. Chemical enrichment happens via three different channels: asymptotic giant branch stars (AGB-stars), Type Ia supernovae and Type II (core-collapse) supernovae including the stellar winds from their progenitors. The nucleosynthetic yield for intermediate-mass AGB-stars are taken from \citet{Marigo2001}, for SNe Ia from \citet{Thielemann2003}, and for mass loss and SNe II in high-mass stars (up to $100\,M_{\odot}$) from \citet{Portinari1998}. These (metallicity-dependent) yields, together with the mass- and metallicity-dependent lifetimes from \citet{Portinari1998}, define the ejected mass of a SSP at each timestep. Nine elements are tracked individually in the chemical enrichment scheme of EAGLE: H, He, C, N, O, Ne, Mg, Si \& Fe\footnote{For the calculation of the radiative cooling rates, Ca and S are additionally taken into account. Within EAGLE, the abundances of Ca and S are proportionally set to the Si abundance.}. We remark that EAGLE does not track individual isotopes, such that the mass fractions in the star and gas particles can be interpreted as comprising all isotopes of the associated element. The simulation also tracks the total metal mass, using $Y_Z=-(Y_{\mathrm{H}}+Y_{\mathrm{He}})$ for the yield of the metals (\citealt{Wiersma2009b}). The yields of hydrogen and helium are negative since destruction dominates over production in the nucleosynthesis process. Specifically relevant for this work is the implementation of the delay time distribution of SN Ia. As described in \citet{Schaye2015} the SN Ia rate of a single star particle follows an exponential decay. The normalisation and e-folding time of 2 Gyr are chosen to reproduce the observed evolution of the cosmic SN Ia rate.

EAGLE uses abundances smoothed by the SPH-kernel (smoothed abundances) for the calculation of cooling rates, stellar lifetimes and yields. Compared to the standard `particle abundances', smoothed abundances have the advantage of being more consistent with the SPH-formalism, furthermore they counter the lack of metal mixing which is inherent to SPH-simulations (\citealt{Wiersma2009b}). Following \citet{Matthee2018} we use smoothed abundances throughout this work.

We note that the simulations we analyse are all assuming a constant Chabrier (\citealt{Chabrier2003}) initial mass function (IMF). In principle, variations in $\alpha$-enhancements could also trace variations of the IMF between different star-forming regions (as for example implemented in \citealt{Barber2019}). Such variations could be relatively independent of variations in star formation histories. For example, a top-heavy IMF will lead to a higher fraction of Type II SNe compared to Type Ia SN, and thus a higher $\alpha$-enhancement \citep{Fontanot2017}. We therefore caution that IMF variations may complicate the applicability of our results to the observed Universe in case IMF variations are occurring.

\subsection{Galaxy sample}\label{sec:GalaxySample}

The RefL0100N1504-simulation consists of 40312 galaxies at redshift zero with stellar masses ranging from $M_{*} \approx10^8-2\times10^{12}\,\mathrm{M}_{\odot}$. The distribution of these galaxies in the stellar mass-sSFR plane (the galactic main sequence) is shown in Figure~\ref{fig:MainSequence}. To minimise resolution effects we do not include low-mass galaxies in our sample as these are resolved by less than 100 star particles. In the simulation, the most massive galaxies tend to be dominated by $\alpha$-enhanced stars and are typically quenched or about to (see Figure~\ref{fig:MainSequence}). Therefore, their star formation histories may have been significantly affected by AGN feedback \citep[e.g.][]{Segers2016, Matthee2019}. To minimise the mass dependence of $\alpha$-enhancement and to reduce the impact of AGN feedback we focus on galaxies in an intermediate mass range and select galaxies within $M_*=(5-10) \times10^9\,\mathrm{M}_{\odot}$, leading to a sample of 1949 galaxies. However, we note that Figure~\ref{fig:MainSequence} illustrates that, for star-forming galaxies, there are little variations in $\alpha$-enhancement as a function of mass, particularly once galaxies that are experiencing AGN feedback are ignored. We investigate star-forming galaxies in the current study, imposing sSFR$>10^{-11}\,\mathrm{yr}^{-1}$ (1455 galaxies). These cuts are illustrated in Figure~\ref{fig:MainSequence}. Lastly, to avoid additional complexity arising due to the different processes acting on the central and satellite populations, we include only central galaxies for our final star-forming sample (960 galaxies). We remark that as long as galaxies are star-forming, it is expected that the physical processes acting on centrals and satellites are similar. We find that our results hardly change when incorporating satellites in the star-forming sample.

We also define a sample of massive, passive galaxies in the RefL0100N1504-simulation. This sample is chosen to allow comparisons between our methodology and some well-known results for the $\alpha$-enhancement of early-type galaxies. As quenching occurs at $\gtrsim10^{10}\,\mathrm{M}_{\odot}$, we choose galaxies with $M_*>3\times10^{10}\,\mathrm{M}_{\odot}$. To ensure that our sample consists of early-type galaxies we only select central galaxies that are devoid of star-forming gas. This passive sample consists of 27 galaxies, the attributes of both samples are summarised in Table~\ref{tab:GalaxySample}. 

Individual star particles in EAGLE, predominantly old ones, can have extremely low or even zero-valued metal mass fractions. Since we build abundance ratios of star particles for the calculation of $[\alpha/\mathrm{Fe}]_{\mathrm{mean}}$ (Eqn.~\ref{eq:alpha_mean}) we have to either assign floor abundances or exclude them from our calculations. We choose the latter option, neglecting all star particles that have Fe, O, Mg or total metal mass fraction below 1/1000 of the corresponding solar value. By varying the cut between $10^{-2}$ and $10^{-5}$ we verified that the choice of this cut does not affect our results in any way.

\begin{table}
    \centering
    \caption{Summary of the properties of the simulated galaxies in the EAGLE simulation at $z=0$ that are used in this work.}
    \begin{tabular}{c|p{5.1cm}|c}
        Sample& Selection & $\mathrm{N_{Gal}}$ \\ \hline
        
        SFGs & $M_*=(5-10) \times10^9\,\mathrm{M}_{\odot}$, $\mathrm{sSFR}>10^{-11}$ yr$^{-1}$, central galaxy & 960 \\
        Passive & $M_*>3 \times10^{10}\,\mathrm{M}_{\odot}$, $\mathrm{SFR}=0\,\mathrm{M}_{\odot}\, \mathrm{yr}^{-1}$, central galaxy & 27\\

    \end{tabular}
    \label{tab:GalaxySample}
\end{table}

\begin{figure}
	\includegraphics[width=\columnwidth]{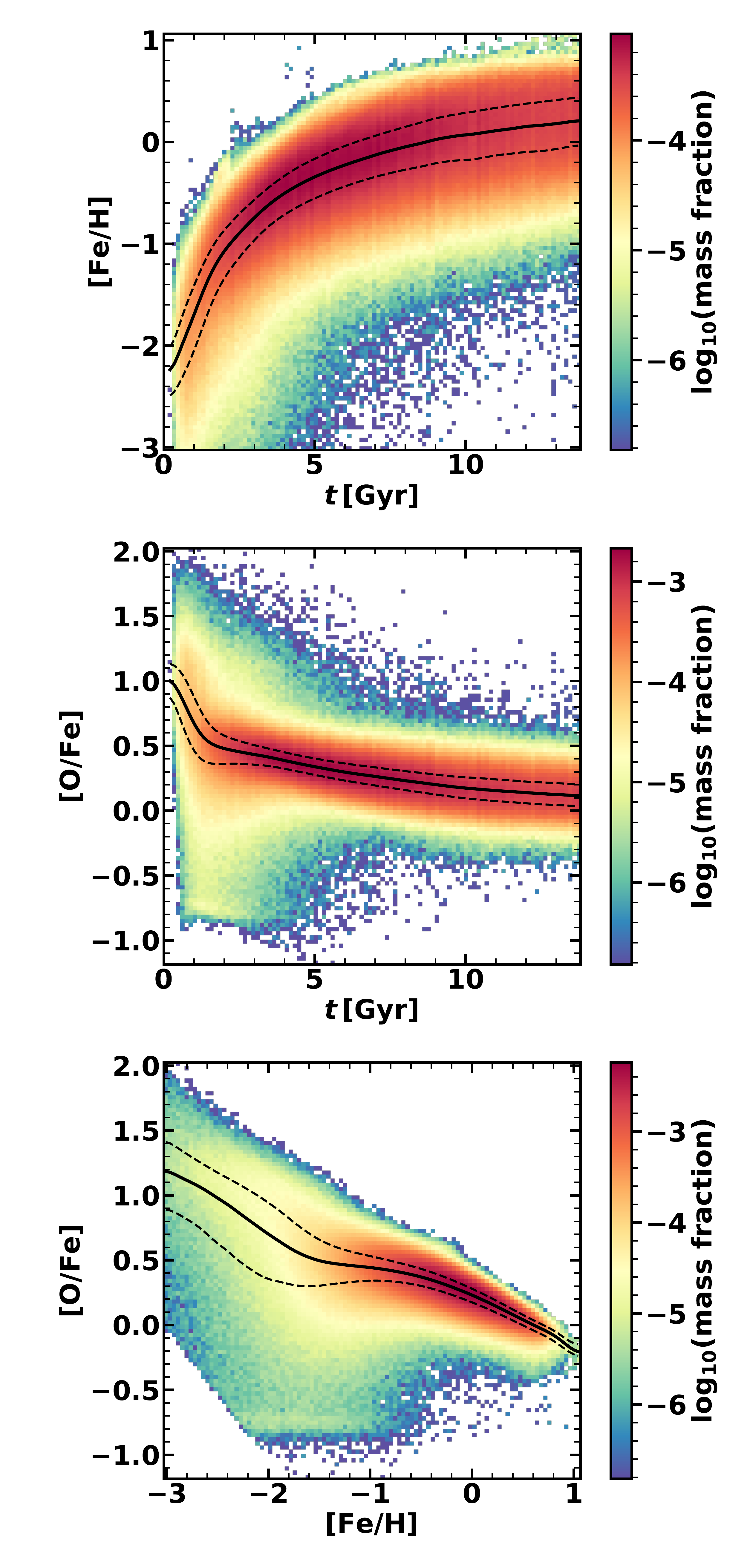}
    \caption{Distribution of the current masses of star particles within our sample of central star-forming galaxies with $M_{*}=(5-10)\times10^9$ M$_{\odot}$ (see \S~\ref{sec:GalaxySample}), in terms of stellar age, metallicity, and $\alpha$-enhancement (linearly discretised to a $100\times100$ grid). $t$ denotes cosmic time, [O/Fe] (and equivalently [Fe/H]) the logarithmic ratio of the oxygen to iron mass fractions (normalised to the solar value). The solid black lines indicate the running median stellar metallicity (top panel) or $\alpha$-enhancement (middle and bottom panels), the dashed lines show the interquartile range (all weighted by current stellar masses).}
    \label{fig:StarParticles}
\end{figure}

\subsection{Census of the Metallicity and $\alpha$-enhancement of star particles in EAGLE}\label{sec:Census}

Since oxygen dominates the mass budget of the $\alpha$-elements, we use oxygen as a proxy for the $\alpha$-elements as done in other studies of the $\alpha$-enhancement in EAGLE (e.g. \citealt{Segers2016}). We remark that while the gas-phase oxygen abundance can be measured reliably from nebular emission lines, the stellar oxygen abundance is difficult to retrieve due to the lack of strong oxygen absorption lines in stellar atmospheres. Hence, some authors opt to use magnesium as a proxy for the $\alpha$-elements. We verified that our results do not change when using magnesium instead of oxygen.

In order to obtain a census picture of the chemical enrichment of all the stars (i.e. star particles representing SSPs) in the galaxies in our star-forming sample, we show the relations between metallicity, oxygen-to-iron ratio and age in Figure~\ref{fig:StarParticles}. The top panel displays the enrichment in iron over time, showing a steep increase in the first $\approx2\,$Gyr followed by continuously slower iron buildup. For the chemical enrichment model implemented in EAGLE, the contribution to the ejected iron mass  of an SSP from SNe Ia becomes relevant only for SSP ages larger than $\approx1\,$Gyr and saturates at $30-45\,\%$ (see Figure 2 in \citealt{Wiersma2009b}). Hence, the rapid initial buildup of iron is dominated by SNe II with a continuously increasing contribution of SNe Ia. The chemical enrichment history is a consequence of the interplay between the SFH, yields and gas cycling effects such that the behaviour in Figure~\ref{fig:StarParticles} does not link trivially to the SSP yields directly. The middle panel shows that after the onset of star formation, [O/Fe] of newborn stars quickly decreases until $t\approx1.5\,$Gyr, after which the decrease becomes more gentle. This behaviour qualitatively agrees with chemical evolution models (e.g. \citealt{Tinsley1979}; \citealt{Sharma2020}; \citealt{Lian2020b}) and other cosmological simulations (e.g. the zoom simulation on a Milky Way analogue VINTERGATAN, \citealt{Agertz2021}). 

The bottom panel of Figure~\ref{fig:StarParticles} shows that for low metallicities $-1.5<\mathrm{[Fe/H]}<-0.5$, the stars are constantly $\alpha$-enhanced, while for higher metallicities [O/Fe] bends over and starts to decrease. This decrease in [O/Fe] is driven by the delayed iron enrichment from SNe Ia, as expected within `time-delay' models of chemical evolution (\citealt{Matteucci2001}; \citealt{Matteucci2012}). Several studies find a similar distribution of stars in the metallicity-$\alpha$-enhancement plane using chemical evolution models (e.g. \citealt{Matteucci1990}; \citealt{Andrews2017}) or cosmological simulations (e.g. \citealt{Mackereth2018}; \citealt{Agertz2021}). On the other hand, surveys of stars in the Milky Way such as APOGEE (\citealt{Majewski2017}) find a bimodal distribution in this plane (e.g. \citealt{Fuhrmann1998}; \citealt{Lian2020b}; \citealt{Lian2020a}). Chemical evolution models of the Galaxy can reproduce this distribution as well as other observational metrics like abundance gradients in the context of the `two-infall' model,
\textbf{where the Galactic thick and thin disks are formed in two gas infall phases initiated at different times and accretion rates (\citealt{Chiappini1997}; \citealt{Fenner2003}; \citealt{Grisoni2017}; \citealt{Spitoni2019})}. \citet{Mackereth2018} use the EAGLE simulation to show that the observed bimodality is indeed only achieved for galaxies with infall periods at both early and late times. Generally, the observed abundance patterns of the Galaxy are attributed to an irregular and atypical gas accretion (and, consequently, star formation) history (\citealt{Mackereth2018}; \citealt{Evans2020}).

\subsection{Alpha-Enhancements of galaxies in EAGLE}\label{sec:Definitions}
Here we apply the general [$\alpha$/Fe] considerations discussed in \S~\ref{sec:Caution} to our analysis of galaxies in the EAGLE simulation. The first definition of the average $\alpha$-enhancement of a galaxy that we consider is the (mass-weighted) average ratio of the stellar oxygen-to-iron abundances of the individual star particles. This corresponds to the `mean $\alpha$-enhancement' which was defined in Eqn.~\ref{eq:alpha_mean}:

\begin{equation}
\begin{split}
    [\mathrm{O/Fe}]_{\mathrm{mean}}&=\log_{10}\Bigg(\bigg\langle \frac{X^{\mathrm{O}}}{ X^{\mathrm{Fe}}}\bigg\rangle_m\Bigg)-\log_{10}\Biggl(\frac{X^{\mathrm{O}}_{\odot}}{X^{\mathrm{Fe}}_{\odot}}\Biggr)\\
    &=\log_{10}\Bigg(\frac{\sum_i m_i\times X_i^{\mathrm{O}}/X_i^{\mathrm{Fe}}}{M_*}\Bigg)-\log_{10}\Biggl(\frac{X^{\mathrm{O}}_{\odot}}{X^{\mathrm{Fe}}_{\odot}}\Biggr),
    \label{eq:OFe_mean}
\end{split}
\end{equation}

where the galactic stellar mass is $M_*=\sum_i m_i$ and $m_i$ are the current masses of all star particles bound to this galaxy. The second definition of the average $\alpha$-enhancement of a galaxy that we consider here is the ratio of the (mass-weighted) average oxygen and iron abundances. This is equivalent to the `galactic $\alpha$-enhancement', defined in Eqn.~\ref{eq:alpha_gal}:

\begin{equation}
\begin{split}
    [\mathrm{O/Fe}]_{\mathrm{gal}}&=\log_{10}\Bigg(\frac{\big\langle X^{\mathrm{O}} \big\rangle_m}{\big\langle X^{\mathrm{Fe}} \big\rangle_m}\Bigg)-\log_{10}\Biggl(\frac{X^{\mathrm{O}}_{\odot}}{X^{\mathrm{Fe}}_{\odot}}\Biggr)\\
    &=\log_{10}\Bigg(\frac{\sum_i m_i\times X_i^{\mathrm{O}}}{\sum_i m_i\times X_i^{\mathrm{Fe}}}\Bigg)-\log_{10}\Biggl(\frac{X^{\mathrm{O}}_{\odot}}{X^{\mathrm{Fe}}_{\odot}}\Biggr).
    \label{eq:OFe_gal}
\end{split}
\end{equation}

Since the galactic mass fraction in element $j$ is given by $X^j_*=\sum_i (m_i\times X^j_i) /M_*$, we can rewrite Eqn.~\ref{eq:OFe_gal} in terms of the galactic oxygen and iron abundances:

\begin{equation}
    [\mathrm{O/Fe}]_{\mathrm{gal}}=\log_{10}\Biggl(\frac{X^{\mathrm{O}}_*}{X^{\mathrm{Fe}}_*}\Biggr)-\log_{10}\Biggl(\frac{X^{\mathrm{O}}_{\odot}}{X^{\mathrm{Fe}}_{\odot}}\Biggr).
    \label{eq:Alpha_gal_a}
\end{equation}

Earlier analyses of $\alpha$-enhancement within EAGLE \citep[][]{Segers2016, Matthee2018, Mackereth2018, Hughes2020} consider exactly this quantity. For the remainder of this paper, we will use [O/Fe]$_{\mathrm{gal}}$ as defined in Eqn.~\ref{eq:OFe_gal} using the EAGLE star particle catalogue, since this definition incorporates an averaging procedure which will make it easier to compare our definitions of `mean' and `galactic' $\alpha$-enhancement.

\begin{figure}
	\includegraphics[width=\columnwidth]{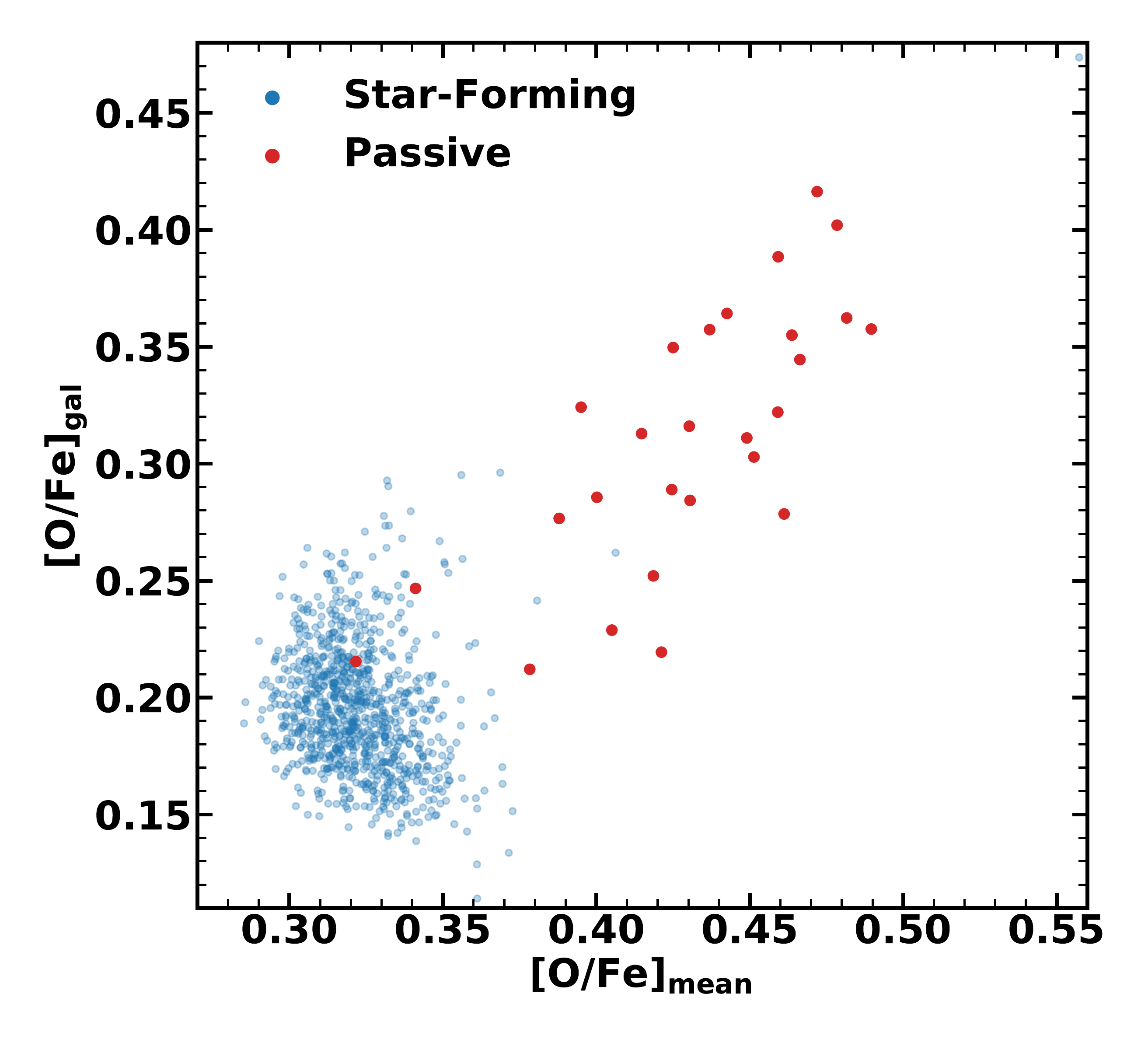}
    \caption{Alpha-enhancements of central EAGLE galaxies at $z=0$, for the star-forming (blue) and passive (red) galaxy samples (Table~\ref{tab:GalaxySample}).}
    \label{fig:AlphaPlane}
\end{figure}

For a SSP (or a CSP where all stars exhibit the same oxygen-to-iron ratio), [O/Fe]$_{\mathrm{mean}}$ and [O/Fe]$_{\mathrm{gal}}$ must coincide. Since the simulated galaxies consist of star particles with varying oxygen-to-iron ratios, we naively expect a strong, positive correlation between [O/Fe]$_{\mathrm{mean}}$ and [O/Fe]$_{\mathrm{gal}}$. We show a comparison of these two definitions of $\alpha$-enhancement for our star-forming and passive galaxy samples in Figure~\ref{fig:AlphaPlane}. We find that the $\alpha$-enhancements of the passive sample indeed correlate quite well with a Pearson correlation coefficient of $\rho=0.75$. This positive correlation among passive galaxies extends to lower masses, a sample of 77 central galaxies with $M_*=(5-10)\times10^9\,\mathrm{M}_{\odot}$ and sSFR$<10^{-11}\,\mathrm{yr}^{-1}$ is distributed similarly ($\rho=0.75$) as the sample of 27 massive, passive centrals (besides the $\alpha$-enhancements of the low-mass passive galaxies being $\approx0.09$\,dex lower on average due to galactic downsizing). On the other hand, [O/Fe]$_{\mathrm{mean}}$ and [O/Fe]$_{\mathrm{gal}}$ are uncorrelated for the star-forming sample ($\rho=-0.02$). This discrepancy between passive and star-forming galaxies is due to the passive galaxies being more similar to SSPs, for which the different $\alpha$-enhancement definitions are equivalent.

We also note that [O/Fe]$_{\mathrm{mean}}$ is significantly larger than [O/Fe]$_{\mathrm{gal}}$, on average the shift is $\approx0.13$ dex. This can be understood in terms of the `metallicity-weighting' of [O/Fe]$_{\mathrm{gal}}$, revealed by rewriting the ratio of the averaged abundances:

\begin{equation}
\begin{split}
    \frac{\big\langle X^{\mathrm{O}}\big\rangle_m}{\big\langle X^{\mathrm{Fe}} \big\rangle_m}&=\frac{\sum_iX_i^{\mathrm{O}}m_i}{\sum_iX_i^{\mathrm{Fe}}m_i}\\&=\frac{\sum_iX_i^{\mathrm{O}}/X_i^{\mathrm{Fe}}\times X_i^{\mathrm{Fe}}m_i}{\sum_iX_i^{\mathrm{Fe}}m_i}=\bigg\langle \frac{X^{\mathrm{O}}}{ X^{\mathrm{Fe}}}\bigg\rangle_{m^{\mathrm{Fe}}}.
    \label{eq:Shuffle}
\end{split}
\end{equation}

Here $m^{\mathrm{Fe}}=X^{\mathrm{Fe}}\times m$ denotes the current iron mass in the star particles. Combining Eqns.~\ref{eq:OFe_gal} and~\ref{eq:Shuffle} leads to

\begin{equation}
    [\mathrm{O/Fe}]_{\mathrm{gal}}=\log_{10}\Bigg(\bigg\langle \frac{X^{\mathrm{O}}}{ X^{\mathrm{Fe}}}\bigg\rangle_{m^{\mathrm{Fe}}}\Bigg)-\log_{10}\Biggl(\frac{X^{\mathrm{O}}_{\odot}}{X^{\mathrm{Fe}}_{\odot}}\Biggr).
    \label{eq:Alpha_gal_c}
\end{equation}

Hence, [O/Fe]$_{\mathrm{mean}}$ and [O/Fe]$_{\mathrm{gal}}$ describe the same quantity, the stellar oxygen-to-iron ratio, but averaged (weighted) by different functions. Note that [O/Fe]$_{\mathrm{mean}}$ uses the intuitive mass-weighting of the stars, while [O/Fe]$_{\mathrm{gal}}$ is weighted by the stellar iron mass. Since iron-rich stars are comparatively young, and young stars are suppressed in $\alpha$-elements (Figure \ref{fig:StarParticles}), [O/Fe]$_{\mathrm{gal}}$ is dominated by young stars and systematically lower than [O/Fe]$_{\mathrm{mean}}$. We remark that the range of $\alpha$-enhancements spanned by the star-forming galaxies in our sample is rather small: the two-$\sigma$-interval for [O/Fe]$_{\mathrm{mean}}$ ranges from $0.30-0.36$ ($\approx0.06$\,dex), the one for [O/Fe]$_{\mathrm{gal}}$ from $0.15-0.26$ ($\approx0.11$\,dex).

\begin{figure*}
	\includegraphics[width=\textwidth]{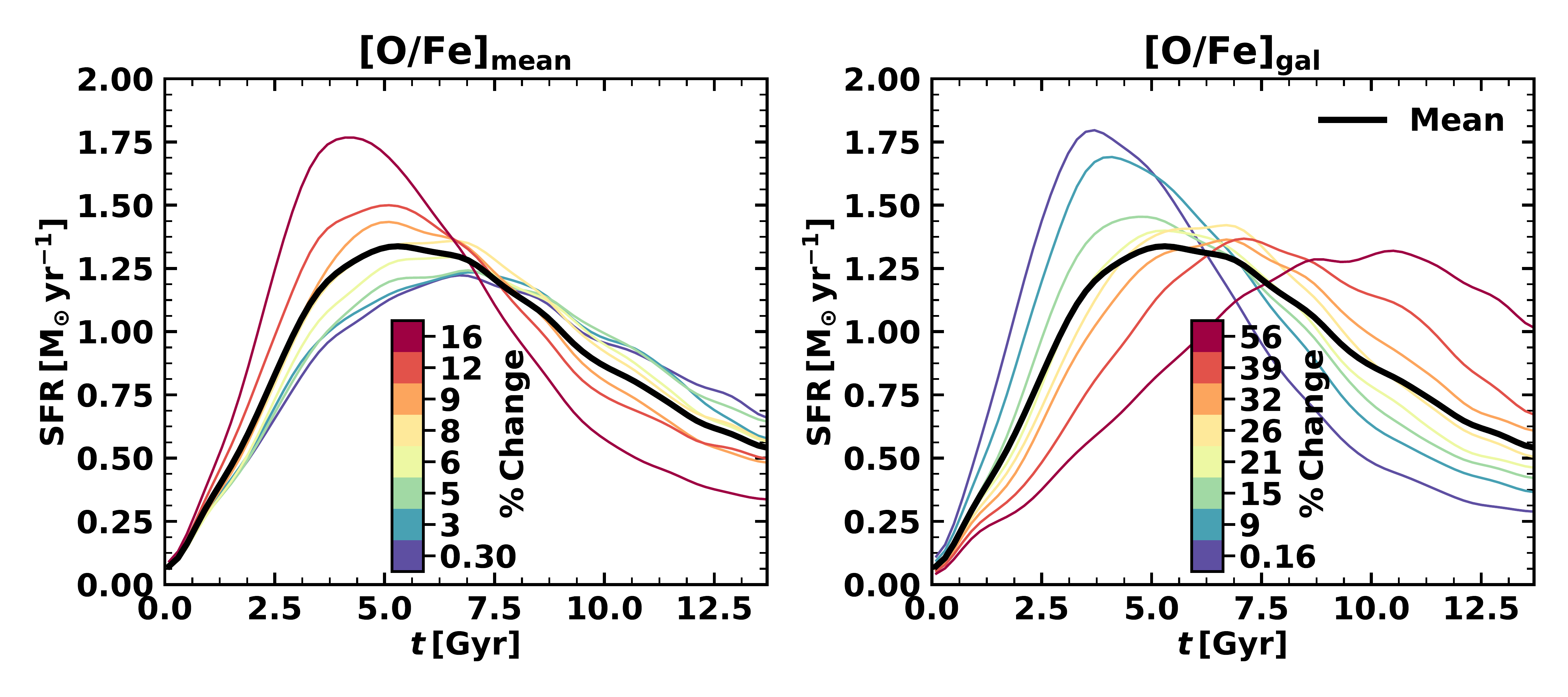}
    \caption{SFHs of star-forming EAGLE galaxies, with $t$ denoting cosmic time. The mean SFH is calculated as the average SFR of all star-forming galaxies in the sample in each time bin of 0.2\,Gyr. For the other curves, the galaxies are split into eight equally-large subsets, sorted by [O/Fe]$_{\mathrm{mean}}$ or [O/Fe]$_{\mathrm{gal}}$. We show the average SFR for each of these subsets. The color scale indicates the level of $\alpha$-enhancement of each subset relative to the subset with the lowest value (which is 0.30 for [O/Fe]$_{\mathrm{mean}}$ and 0.16 for [O/Fe]$_{\mathrm{gal}}$) in percent. There is significantly more scatter in [O/Fe]$_{\mathrm{gal}}$ compared to [O/Fe]$_{\mathrm{mean}}$.}
    \label{fig:SFHs}
\end{figure*}

\section{Connection between Star Formation Histories and $\alpha$-enhancement}\label{sec:Evolution}

In this section we focus on the connection between the two definitions of $\alpha$-enhancement and the star formation history. Due to the different timescales associated to the enrichment of $\alpha$ elements and iron, a connection between $\alpha$-enhancement and SFH is expected \citep[e.g.][]{Thomas2005,Rosa2011}. However, we have seen that the two $\alpha$-enhancements anti-correlate slightly for star-forming galaxies, suggesting that the single interpretation of $\alpha$-enhancement as a tracer for the rapidness of the SFH is over-simplistic.

Starting from our definitions of $\alpha$-enhancement, we can expand on Eqn.~\ref{eq:OFe_mean} and reveal a \textit{twofold dependency} of [O/Fe]$_{\mathrm{mean}}$ on the SFH:

\begin{equation}
\begin{split}
    [\mathrm{O/Fe}]_{\mathrm{mean}}&=\log_{10}\Bigg(\frac{1}{M_*}\int\Big(\frac{X^{\mathrm{O}}_{\mathrm{sf}}}{ X^{\mathrm{Fe}}_{\mathrm{sf}}}\Big)(t)\frac{\mathrm{d}M_*}{\mathrm{d}t}\,\mathrm{d}t\Bigg)\\&-\log_{10}\Biggl(\frac{X^{\mathrm{O}}_{\odot}}{X^{\mathrm{Fe}}_{\odot}}\Biggr),
    \label{eq:Alpha_mean_t}
\end{split}
\end{equation}

where $X^{\mathrm{Fe}}_{\mathrm{sf}}/X^{\mathrm{Fe}}_{\mathrm{sf}}(t)$ is the average\footnote{If multiple star particles are formed between $t$ and $t+\mathrm{d}t$, we weigh the abundance ratios by the current (at redshift zero) stellar mass of the particle to calculate the average abundance ratio.} oxygen-to-iron ratio in star particles newly formed between $t$ and $t+\mathrm{d}t$. Since newborn star particles inherit the abundances from their parent gas particles in EAGLE, $X^{\mathrm{Fe}}_{\mathrm{sf}}/X^{\mathrm{Fe}}_{\mathrm{sf}}(t)$ is essentially the evolving gas-phase oxygen-to-iron ratio of the galaxy. $\mathrm{d}M_*/\mathrm{d}t$ denotes the \textit{current} stellar mass formed and assembled between $t$ and $t+\mathrm{d}t$ as function of cosmic time $t$. This quantity closely resembles the definition of the star formation history\footnote{The star formation history is defined as the function SFR$(t)$, which is the \textit{initial} mass formed per unit time.}, and equals the actual galactic SFH when neglecting stellar mass loss. Since the gas-phase oxygen-to-iron ratio depends on the past star formation history, the SFH affects [O/Fe]$_{\mathrm{mean}}$ both directly (as `weighting' function $\mathrm{d}M_*/\mathrm{d}t$) and indirectly (via $X^{\mathrm{O}}_{\mathrm{sf}}/X^{\mathrm{Fe}}_{\mathrm{sf}}(t)$).

Mathematically, [O/Fe]$_{\mathrm{mean}}$ is fully determined by the oxygen-to-iron ratio of newborn stars and the current mass formed per unit time (both functions of cosmic time $t$). For [O/Fe]$_{\mathrm{gal}}$ we need the current \textit{iron} mass formed per unit time to weigh the stellar oxygen-to-iron ratio, i.e. $\mathrm{d}M^{\mathrm{Fe}}_{*}/\mathrm{d}t$ Hence, with a set of three functions of time one can calculate both [O/Fe]$_{\mathrm{mean}}$ and [O/Fe]$_{\mathrm{gal}}$. Note that this \textit{basis set} is not unique, to facilitate the physical interpretation of these functions we analyse the following three time-dependent quantities in \S~\ref{sec:SFHs} and~\ref{sec:chemevo}: SFR$(t)$, $X^{\mathrm{O}}_{\mathrm{sf}}/X^{\mathrm{Fe}}_{\mathrm{sf}}(t)$, and $X^{\mathrm{Fe}}_{\mathrm{sf}}(t)$.

\subsection{Star Formation Histories}\label{sec:SFHs}

We calculate SFHs using a similar procedure as \citet{Sparre2015}. The star particles of a particular galaxy are binned into age bins of 0.2\,Gyr. Then, we sum the \textit{initial} masses of all star particles formed within the individual age bins. This sum, divided by 0.2\,Gyr, yields the SFR as a function of time. As noted by \citet{Diemer2017}, this extraction of the SFR means that we do not differ between stars formed in situ (within the galaxy in question, which we would for example do when following the merger trees of the simulated galaxies) or ex situ (in a smaller galaxy that merged within the time interval). We proceed by grouping the galaxies into eight equally-large subsets, sorted by [O/Fe]$_{\mathrm{mean}}$ or [O/Fe]$_{\mathrm{gal}}$, and calculate average SFHs for each of the subsets. These average SFHs are shown in Figure~\ref{fig:SFHs}, which also displays the average SFH for the entire star-forming sample of 960 galaxies as a black line in both panels. 

Remarkably, the SFH of the highest [O/Fe]$_{\mathrm{mean}}$ bin coincides rather well with the SFH of the bin with the lowest [O/Fe]$_{\mathrm{gal}}$. These SFHs peak early and have rapid star-formation over a comparatively small time-interval. Such rapid SFHs are expected to have a high $\alpha$-enhancement (e.g. \citealt{Thomas2005}), since the bulk of the stars formed early from an ISM which is not yet iron-enriched by SNe Ia (see also Figure~\ref{fig:StarParticles}). We emphasize that while [O/Fe]$_{\mathrm{mean}}$ follows this theoretical picture, [O/Fe]$_{\mathrm{gal}}$ behaves exactly in the opposite way. At first sight, this is a puzzling behaviour. However, as shown in Eqn.~\ref{eq:Alpha_gal_c}, [O/Fe]$_{\mathrm{gal}}$ is in fact the metallicity-weighted version of [O/Fe]$_{\mathrm{mean}}$. As the metallicity evolves by up to three orders of magnitude over the lifetime of a galaxy (see Figure~\ref{fig:StarParticles}), this means that [O/Fe]$_{\mathrm{gal}}$ is mostly determined by the relatively late-time shape of the SFH. As we discuss in detail in \S~\ref{sec:chemevo}, this dominant late-time impact together with the chemical evolution leads to the observed correlation between the SFH and [O/Fe]$_{\mathrm{gal}}$.

Figure \ref{fig:SFHs} also shows that at the low-[O/Fe]$_{\mathrm{mean}}$-end the differences in the late SFHs become smaller, indicating a saturation of the effect that SFH parameters correlate with $\alpha$-enhancement. This means that it is mostly early star formation that determines [O/Fe]$_{\mathrm{mean}}$. Contrarily, for [O/Fe]$_{\mathrm{gal}}$ we find significant differences both at early and late times. The continuous ordering of the SFHs indicates a monotonic correlation between the shape of the SFH and the $\alpha$-enhancements. Furthermore, we notice that the average SFHs have a log-normal shape, as is found in other studies for simulated galaxies (e.g. \citealt{Diemer2017}). We fit log-normal functions to the SFHs of the individual galaxies in our sample in \S~\ref{sec:Fitting SFHs}.

\begin{figure*}
	\includegraphics[width=0.92\textwidth]{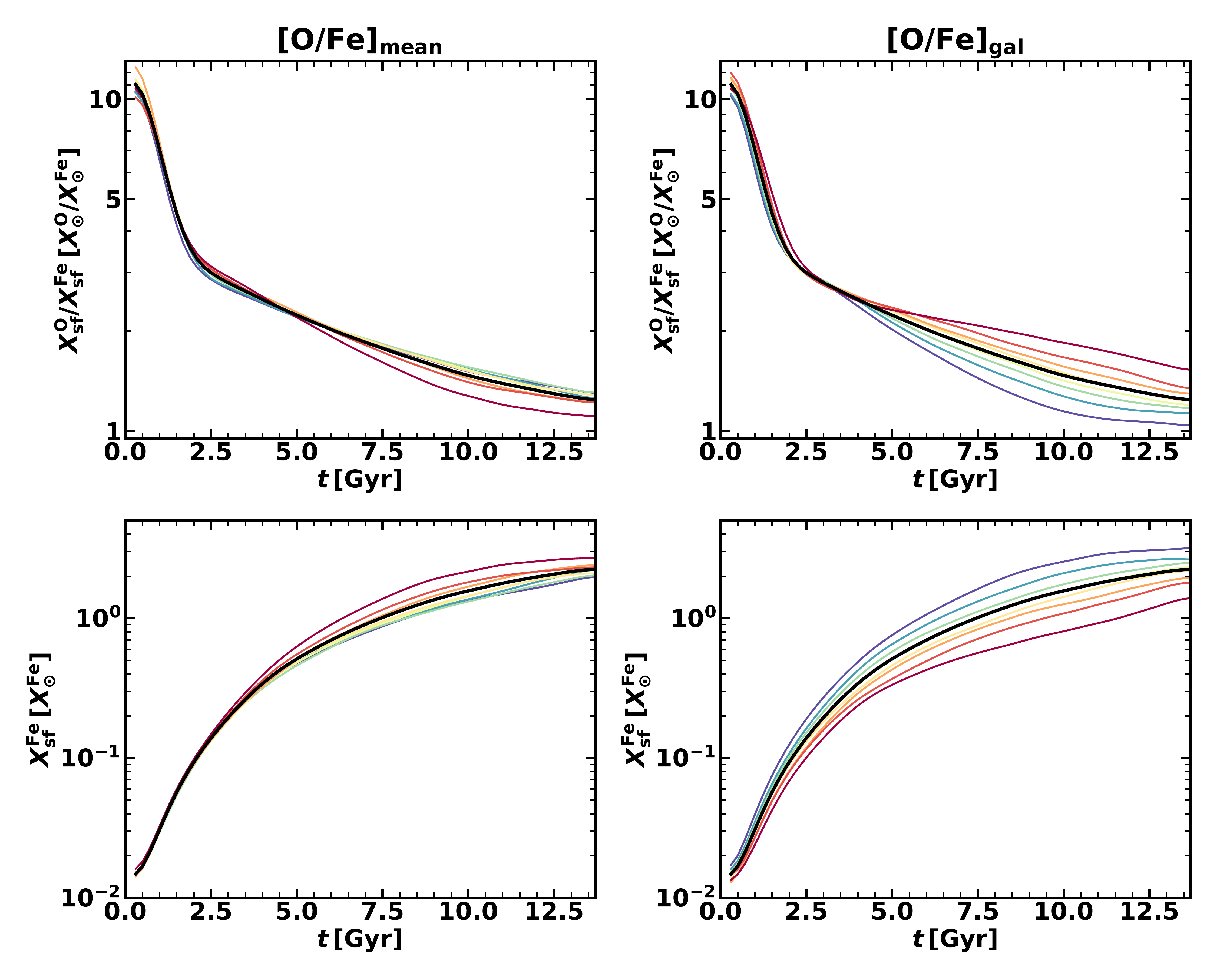}
    \caption{Chemical evolution of star-forming EAGLE galaxies parametrised by the abundances of newborn stars, equivalent color-coding as in Figure \ref{fig:SFHs}. The mean $X^{\mathrm{O}}_{\mathrm{sf}}/X^{\mathrm{Fe}}_{\mathrm{sf}}$ and  $X^{\mathrm{Fe}}_{\mathrm{sf}}$ (black curves) are averaged over all galaxies in the sample in each time bin of 0.2\,Gyr. For the other curves, the galaxies are split into eight equally-sized subsets, sorted by $[\mathrm{O/Fe}]_{\mathrm{mean}}$ or $[\mathrm{O/Fe}]_{\mathrm{gal}}$. We show the average $X^{\mathrm{Fe}}_{\mathrm{sf}}$ and $X^{\mathrm{O}}_{\mathrm{sf}}/X^{\mathrm{Fe}}_{\mathrm{sf}}$ for each of these subsets.}
    \label{fig:CEHs}
\end{figure*}

\begin{figure*}
	\includegraphics[width=0.95\textwidth]{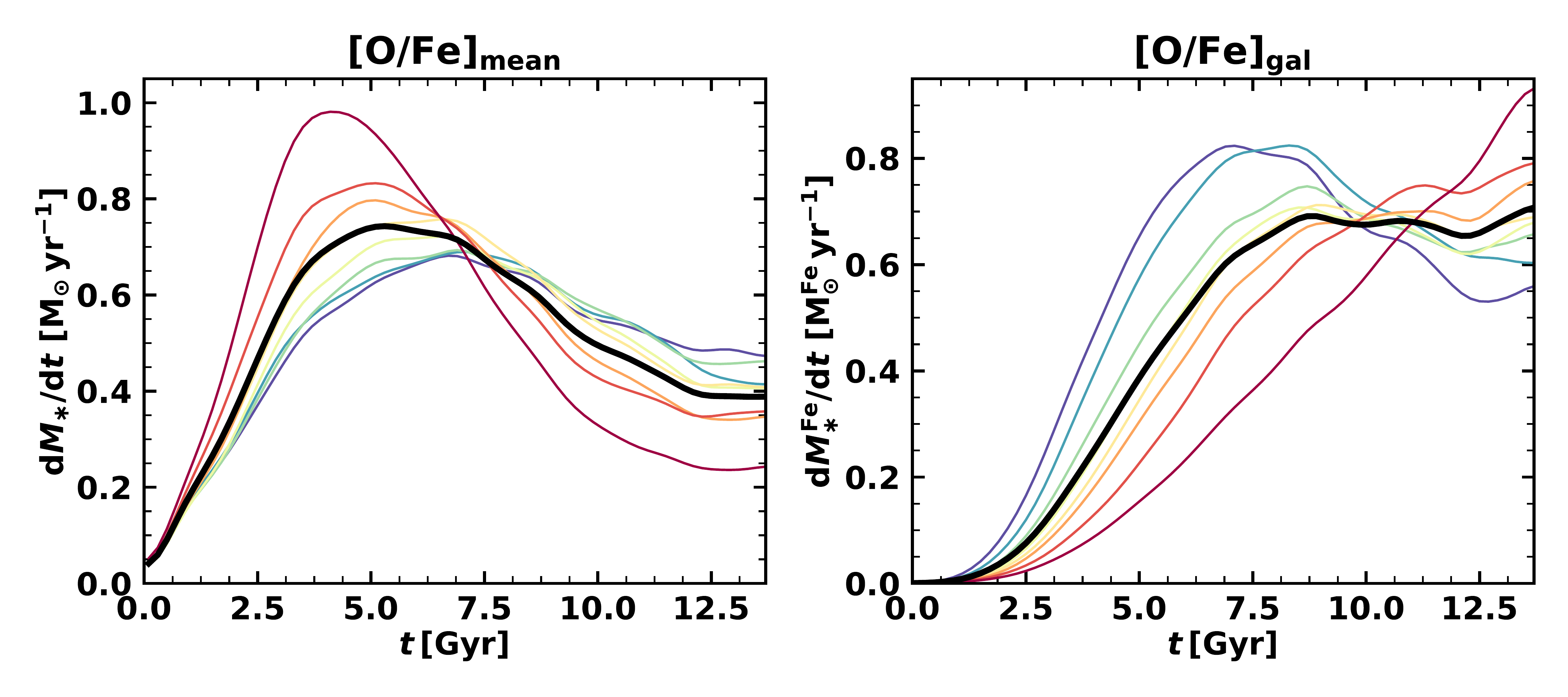}
    \caption{Star formation histories as in Figure~\ref{fig:SFHs}, but with the actual weighting functions used for [O/Fe]$_{\mathrm{mean}}$ (current galactic mass formed per unit time, left panel) and [O/Fe]$_{\mathrm{gal}}$ (current galactic \textit{iron} mass formed per unit time, right panel). The galaxies are split into eight equally-sized subsets according to their [O/Fe]$_{\mathrm{mean}}$ and [O/Fe]$_{\mathrm{gal}}$, respectively. We show the average weighting functions within each of these subsets, the black lines indicate the mean tracks of the entire star-forming galaxy sample. The color-coding coincides with the one of Figure~\ref{fig:SFHs}.}
    \label{fig:Weighting Functions}
\end{figure*}

\subsection{Chemical Evolution}\label{sec:chemevo}

We analyse the oxygen-to-iron ratio and the iron abundance of newborn stars as a function of time. Similar as for the SFHs, we bin the star particles into age bins of 0.2\,Gyr, and then compute $X^{\mathrm{O}}_{\mathrm{sf}}/X^{\mathrm{Fe}}_{\mathrm{sf}}\equiv\langle X^{\mathrm{O}}/X^{\mathrm{Fe}}\rangle_m$ and $X^{\mathrm{Fe}}_{\mathrm{sf}}\equiv\langle X^{\mathrm{Fe}}\rangle_m$ for the star particles born within each age bin. The results for galaxies grouped by their $\alpha$-enhancements are shown in Figure~\ref{fig:CEHs}. Differences between the curves arise mostly at late times. For both definitions of $\alpha$-enhancement, the oxygen-to-iron ratios imitate the SFR at late times: a high SFR today leads to a high $X^{\mathrm{O}}_{\mathrm{sf}}/X^{\mathrm{Fe}}_{\mathrm{sf}}$ today and vice versa. This is a consequence of the correlation between SFH and chemical enrichment: galaxies with an early-peaked SFH (hence small SFR today) have their ISM enriched in iron such that stars that form at late times have a small oxygen-to-iron ratio. Comparing the upper panels with the lower panels in Figure~\ref{fig:CEHs}, we find that the curves appear to be mirrored. This behaviour indicates that star particles with a small (large) oxygen-to-iron ratio have this value mostly due to a large (small) iron abundance. As expected, the iron abundance is a strongly increasing function of time (more than two orders of magnitude over the age of the Universe), and galaxies with delayed star-formation (late-peaked SFHs) have smaller iron abundances throughout.

With these two functions - SFR and $X^{\mathrm{O}}_{\mathrm{sf}}/X^{\mathrm{Fe}}_{\mathrm{sf}}$ - we can explain how a galaxy ends up with a certain mean $\alpha$-enhancement today. Galaxies with high [O/Fe]$_{\mathrm{mean}}$ have an early-peaked SFH, such that [O/Fe]$_{\mathrm{mean}}$ is dominated by old stars. Since these older stars have comparatively large oxygen-to-iron ratios, the galaxy ends up with a large value of [O/Fe]$_{\mathrm{mean}}$. Note that these galaxies have a subnormal oxygen-to-iron ratio at late times (Figure~\ref{fig:CEHs}), but this does not affect the galactic [O/Fe]$_{\mathrm{mean}}$ significantly as these galaxies consist of only few young stars (Figure~\ref{fig:SFHs}). Galaxies with a low [O/Fe]$_{\mathrm{mean}}$ have a subnormal star formation activity at early ($t<7\,\mathrm{Gyr}$) times, leading to less young stars with high oxygen-to-iron ratios in these galaxies. Apparently, galaxies with low [O/Fe]$_{\mathrm{mean}}$ do not have excessive star formation today, at least not to the degree as the high-[O/Fe]$_{\mathrm{gal}}$-galaxies (Figure~\ref{fig:SFHs}).

To explain how a galaxy ends up with a certain galactic $\alpha$-enhancement we need to invoke yet another function: the iron abundance of newborn stars, $X^{\mathrm{Fe}}_{\mathrm{sf}}$, as a function of time. While $X^{\mathrm{Fe}}_{\mathrm{sf}}(t)$ does not affect [O/Fe]$_{\mathrm{mean}}$ directly, it does enter the averaging process for the calculation of the galactic $\alpha$-enhancement [O/Fe]$_{\mathrm{gal}}$. Metal-rich star particles receive significantly more weight than metal-poor ones for the computation of the galactic $\alpha$-enhancement. This effect is strong enough to approximately reverse the behaviour of galaxies when sorted by their [O/Fe]$_{\mathrm{gal}}$ (compared to when they are sorted by [O/Fe]$_{\mathrm{mean}}$). For galaxies to achieve a large value of [O/Fe]$_{\mathrm{gal}}$, it is more `efficient' to have late-peaked, extended SFHs (Figure~\ref{fig:SFHs}). This leads to a subnormal amount of old, $\alpha$-enhanced stars, but these are negligible within the averaging process of the galactic $\alpha$-enhancement. The late-peaked SFH ensures that younger stars (formed after $t=7\,\mathrm{Gyr}$), which dominate the value of [O/Fe]$_{\mathrm{gal}}$, are $\alpha$-enhanced compared to other curves at the same age (Figure~\ref{fig:CEHs}). Vice versa, galaxies with a small value of [O/Fe]$_{\mathrm{gal}}$ have early-peaked SFHs and build a lot of iron early on, such that younger stars formed from the iron-rich ISM have comparatively low oxygen-to-iron ratios.

For completeness we also show the actual weighting functions used for mean ($\mathrm{d}M_*/\mathrm{d}t$) and galactic ($\mathrm{d}M_*^{\mathrm{Fe}}/\mathrm{d}t$) $\alpha$-enhancement in Figure~\ref{fig:Weighting Functions}. These functions directly give the $\alpha$-enhancements of the galaxies when connected with the evolving oxygen-to-iron ratios (upper panels of Figure~\ref{fig:CEHs}). The left panel of Figure~\ref{fig:Weighting Functions} shows the current galactic mass formed per unit time, for groups of galaxies split by [O/Fe]$_{\mathrm{mean}}$. The tracks closely resemble the actual star formation histories (left panel of Figure~\ref{fig:SFHs}), scaled down by approximately forty percent which corresponds to the integrated mass loss of an SSP \citep{Wiersma2009b}. This scaling is essentially uniform throughout cosmic history except for the most recent Gyr. Accordingly, the right panel of Figure~\ref{fig:Weighting Functions} shows the current galactic iron mass formed per unit time, for groups of galaxies split by [O/Fe]$_{\mathrm{gal}}$. The tracks are essentially a convolution of the SFHs (right panel in Figure~\ref{fig:SFHs}), the iron abundance histories (lower right panel in Figure~\ref{fig:CEHs}), and the SSP mass loss effects. These tracks show quantitatively that younger stars are dominant for setting [O/Fe]$_{\mathrm{gal}}$, opposed to [O/Fe]$_{\mathrm{mean}}$ which is mostly set by older stars.

\subsection{Fitting Star Formation Histories}\label{sec:Fitting SFHs}

In order to explore the apparent relation between SFHs and the $\alpha$-enhancements of galaxies (Figure~\ref{fig:SFHs}) in a quantitative way and on the level of individual galaxies, we parametrise galactic SFHs and investigate the relation between their fitted parameters and [O/Fe]$_{\mathrm{mean}}$ and [O/Fe]$_{\mathrm{gal}}$. Following \citet{Gladders2013} and \citet{Diemer2017} we describe SFHs of individual galaxies by log-normal functions, given by the following expression:

\begin{equation}\label{eq:Log-Normal}
    \mathrm{SFR}(t)=\frac{A}{\sqrt{2\pi\tau^2}\times t}\exp\Biggl(-\frac{\bigl(\ln(t)-T_0\bigr)^2}{2\tau^2}\Biggr),
\end{equation}

where $A$, $T_0$ and $\tau$ are free parameters (the latter two have units of Gyr). For readability we refrain from inserting unit factors such as $\ln(t/1\,\mathrm{Gyr})$ into the equations in this section, and remark that we measure times always in Gyr and star formation rates in $M_{\odot}\,\mathrm{yr}^{-1}$. We use the open-source \texttt{lmfit} package for \texttt{python} to execute least-squares minimisations with the built-in Levenberg-Marquardt-algorithm, fitting Eqn.~\ref{eq:Log-Normal} to the simulated SFHs. The simulated SFHs are calculated as described in \S~\ref{sec:SFHs} and age bins without any star-formation are ignored by the fitting routine. Due to the discrete nature and implementation of star-formation in the simulation our calculated SFHs contain Poisson noise. Following \citet{Matthee2019} we calculate the associated error as $\sigma(t)=\mathrm{SFR}(t)/\sqrt{N}$, where $N$ is the number of star particles formed in a particular age bin. Hence, the fitting routine effectively corresponds to a minimization of the following quantity:

\begin{equation}\label{eq:Chi-squared}
    \chi^2=\sum_{t=t_i}\Bigl(\frac{\mathrm{SFR}_{\mathrm{sim}}(t)-\mathrm{SFR}_{\mathrm{fit}}(t)}{\sigma(t)}\Bigr)^2,
\end{equation}

where the sum runs over all age bins (excluding bins without star formation activity). Following \citet{Diemer2017}, we replace $T_0$ and $\tau$ with the more convenient peak time of the SFH ($t_{\mathrm{peak}}$, Eqn.~\ref{eq:t_peak}) and FWHM ($\sigma_{\mathrm{SFR}}$, Eqn.~\ref{eq:sigma}):

\begin{equation}\label{eq:t_peak}
    t_{\mathrm{peak}}=\exp\bigl(T_0-\tau^2\bigr).
\end{equation}

\begin{equation}\label{eq:sigma}
    \sigma_{\mathrm{SFR}}=2t_{\mathrm{peak}}\sinh\bigl(\sqrt{2\ln(2)}\tau\bigr).
\end{equation}

\begin{figure}
	\includegraphics[width=\columnwidth]{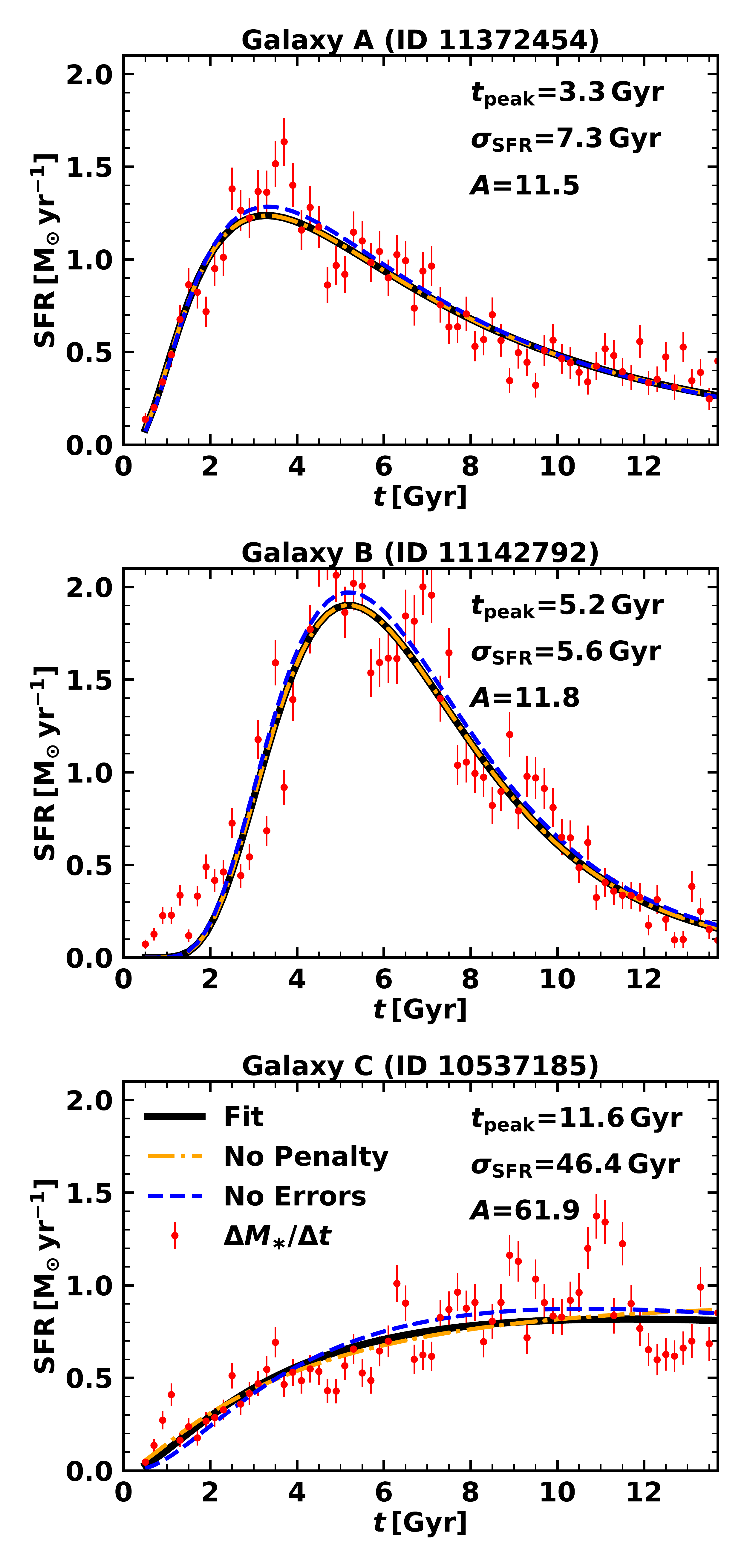}
    \caption{Comparison of the fitted SFHs (black curves) for three star-forming galaxies spanning a large range in $t_{\mathrm{peak}}$ and $\sigma_{\mathrm{SFR}}$. The SFR(t) of simulated galaxies are indicated as red points with the corresponding Poissonian error bars. For comparison we also show the best-fitting log-normal SFHs without using the penalty terms (dash-dotted yellow curves) and without weighing the star formation rates with Poisson noise (dashed blue curves). The fitted parameters which are written in the panels correspond to the regular fitting procedure (black curves).}
    \label{fig:fitted SFH example}
\end{figure}

As noted by \citet{Diemer2017}, a small fraction of galaxies have rising SFHs today, leading to poorly conditioned log-normal fits with unphysically large values of $t_{\mathrm{peak}}$ and $\sigma_{\mathrm{SFR}}$. To address this effect we introduce a penalty term by multiplying the residuals with $1+0.5\bigl(\log_{10}(t_{\mathrm{peak}})-\log_{10}(13.6\,\mathrm{Gyr})\bigr)^2$ if $t_{\mathrm{peak}}>13.6\,\mathrm{Gyr}$. For $\sigma_{\mathrm{SFR}}>20\,\mathrm{Gyr}$ we have a similar penalty term: $1+0.2\bigl(\log_{10}(\sigma_{\mathrm{SFR}})-\log_{10}(20\,\mathrm{Gyr})\bigr)^2$. Although some of the fitted peak times and SFH widths are significantly affected, the fit quality barely changes due to these penalty terms. 65 galaxies (out of the total star-forming sample of 960 objects) have their $\chi^2$ (Eqn.~\ref{eq:Chi-squared}) enlarged by more than one percent, 24 galaxies by more than ten percent, and no galaxy experiences a change in $\chi^2$ larger than sixty percent.

\begin{figure*}
	\includegraphics[width=\textwidth]{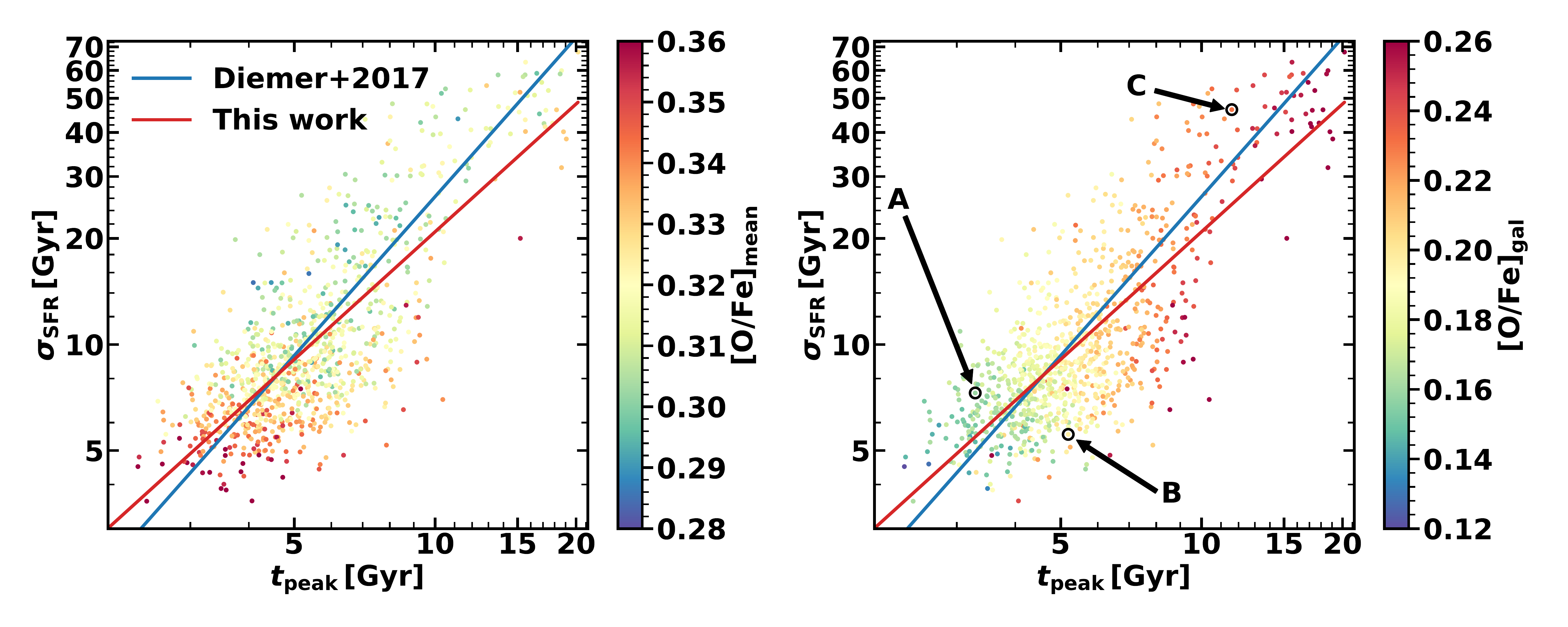}
    \caption{Fitted SFH parameters of the galaxies in our star-forming sample, color-coded by [O/Fe]$_{\mathrm{mean}}$ and [O/Fe]$_{\mathrm{gal}}$. The range of the colorbar is limited and one galaxy with peculiarly small $t_{\mathrm{peak}}$ and $\sigma_{\mathrm{SFR}}$ is excluded from the figure for optimal visualisation. The best-fitting power laws for the $t_{\mathrm{peak}}$-$\sigma_{\mathrm{SFR}}$-relation from \citet{Diemer2017} (blue line) and this work (red line) are also shown. Furthermore, the individual galaxies investigated in Figure~\ref{fig:fitted SFH example} (A: ID 11372454, B: ID 11142792, C: ID 10537185) are marked in the right panel.}
    \label{fig:SFH params and Alpha-Enhancement}
\end{figure*}

Figure~\ref{fig:fitted SFH example} shows fitted SFHs (black curves) for three galaxies spanning a wide range in the $t_{\mathrm{peak}}$-$\sigma_{\mathrm{SFR}}$-parameter space (these galaxies are marked in Figure~\ref{fig:SFH params and Alpha-Enhancement}). The unique galaxy IDs from the EAGLE galaxy catalogue are given in the headers. We illustrate the effect of the penalty terms through the yellow dash-dotted fits, which were calculated without multiplying the penalty terms. Only the SFH in the lowest panel exhibits visible differences between the two SFHs. This is a galaxy with a SFH that increases up to the present day leading to large values of $t_{\mathrm{peak}}$ and $\sigma_{\mathrm{SFR}}$ (even for the regular fit invoking the penalty terms). We also show fits without using the Poisson noise $\sigma(t)$ (dashed blue curves), these fits lie almost always slightly above the regular ones. This is because these fits are dominated by bins with large star formation rates. For a single age bin with a certain SFR, assume that the fitted star formation rate deviates by a factor of $\kappa$. Without accounting for Poisson noise, we then have for this bin $\chi_i^2=(\mathrm{SFR}-\kappa\cdot\mathrm{SFR})^2=(1-\kappa)^2\mathrm{SFR}^2\propto\mathrm{SFR}^2$. Hence, bins with a comparatively large SFR react more sensibly to (multiplicative) deviations of the fitted SFH. This scaling is weakened when invoking the Poisson noise, where $\chi_i^2=(\mathrm{SFR}-\kappa\cdot\mathrm{SFR})^2/(\mathrm{SFR}/\sqrt{N})^2\propto\mathrm{SFR}^2/\sqrt{\mathrm{SFR}}^2\propto\mathrm{SFR}$. The regular fits using Poisson noise are still dominated by bins with large SFRs, but to a lesser extent compared to fitting routines without assigning errors.

The parameter space that galaxy SFHs span in terms of $t_{\mathrm{peak}}$ and $\sigma_{\mathrm{SFR}}$ is shown in Figure~\ref{fig:SFH params and Alpha-Enhancement}. We omit the normalisation $A$ in this discussion as we did not find any relation between stellar mass and $\alpha$-enhancement for the galaxies in our star-forming sample within its narrow mass range. As \citet{Diemer2017} found using the cosmological {\it Illustris} simulation, $t_{\mathrm{peak}}$ and $\sigma_{\mathrm{SFR}}$ correlate approximately linearly in log-log space: $\log_{10}(\sigma_{\mathrm{SFR}})=1.5\log_{10}(t_{\mathrm{peak}})-0.08$ (blue line in Figure~\ref{fig:SFH params and Alpha-Enhancement}). We find a slightly shallower relation of $\log_{10}(\sigma_{\mathrm{SFR}})=1.20\log_{10}(t_{\mathrm{peak}})+0.12$ (red line in Figure~\ref{fig:SFH params and Alpha-Enhancement}). We note that for our log-normal parameter distribution the power law seems to be shallower at early times and becomes steeper for increasing $t_{\mathrm{peak}}$, unlike in \citet{Diemer2017}. Differences could arise due to different galaxy samples (\citealt{Diemer2017} consider all galaxies with $M_*\geq10^9\,M_{\odot}$), different SFHs in EAGLE and {\it Illustris} (see \citealt{Iyer2020} for an in-depth discussion on the differences in SFHs of various cosmological and semi-analytical models) or varying fitting routines (\citealt{Diemer2017} fit the cumulative SFH with an integrated log-normal function, furthermore these authors do not assign Poisson noise).

With the results from the SFH fitting routine we can investigate how individual SFHs correlate with the $\alpha$-enhancements of the galaxies, shown as color-coding in Figure~\ref{fig:SFH params and Alpha-Enhancement}. We note the following points:

\begin{itemize}

    \item Galaxies with high [O/Fe]$_{\mathrm{mean}}$ have small SFH widths, and - as subdominant effect - early peak times.
    
    \item Galaxies with low [O/Fe]$_{\mathrm{mean}}$ are broadly located at the high-$t_{\mathrm{peak}}$, high-$\sigma_{\mathrm{SFR}}$ region.
    
    \item Galaxies with high [O/Fe]$_{\mathrm{gal}}$ have late peak times and - as subdominant effect - small SFH widths. 
    
    \item Galaxies with small [O/Fe]$_{\mathrm{gal}}$ have early peak times and - as subdominant effect - small SFH widths.
    
\end{itemize}

Noting the first and the last point of this list, this explains why the tracks of the high-[O/Fe]$_{\mathrm{mean}}$-galaxies and the low-[O/Fe]$_{\mathrm{gal}}$-ones closely resemble each other in Figure~\ref{fig:SFHs}. In the following, we argue how these observed correlations between the SFH parameters, $\sigma_{\mathrm{SFR}}$ and $t_{\mathrm{peak}}$, and our two definitions of $\alpha$-enhancements of galaxies emerge.

For [O/Fe]$_{\mathrm{mean}}$, small SFH widths correspond to $\alpha$-enhanced galaxies, in line with many studies that explain the high $\alpha$-enhancement of passive galaxies through a fast and rapid SFH (e.g. \citealt{Thomas2005,Arrigoni2010,Segers2016}). The parameter $t_{\mathrm{peak}}$ plays a subdominant role. Small peak times lead to more stars forming from $\alpha$-enhanced gas, but galaxies with large $t_{\mathrm{peak}}$ can quickly enhance the oxygen-to-iron ratio in the ISM by a burst of star formation (i.e. small $\sigma_{\mathrm{SFR}}$) and then form stars out of this $\alpha$-enhanced gas. Reverting this reasoning to explain how a galaxy has low [O/Fe]$_{\mathrm{mean}}$ is apparently not straightforward, as the correlation fades for larger $\sigma_{\mathrm{SFR}}$ and $t_{\mathrm{peak}}$. This weakening of the correlation of SFHs and [O/Fe]$_{\mathrm{mean}}$ for lower values of $\alpha$-enhancement is also visible in Figure~\ref{fig:SFHs}, where the low-[O/Fe]$_{\mathrm{mean}}$-tracks are only weakly separated. We argue that this behaviour is due to $\sigma_{\mathrm{SFR}}$ being the dominant driver of mean $\alpha$-enhancement, and while a large $\sigma_{\mathrm{SFR}}$ leads to stronger iron enrichment of the ISM, the distributed star formation inevitably means that more stars are formed at early times when the ISM is very $\alpha$-enhanced. These two effects oppose each other such that the correlation fades at the low-[O/Fe]$_{\mathrm{mean}}$-end. Crucially, the reasoning that $\sigma_{\mathrm{SFR}}$ has two opposing effects on  [O/Fe]$_{\mathrm{mean}}$ only applies to large peak times, else we would not observe any correlation at the high-[O/Fe]$_{\mathrm{mean}}$-end. This is because for small peak times, an extended SFH does not lead to more stars being formed at early times simply because there is a boundary at $t=0\,$Gyr.

The right panel of Figure~\ref{fig:SFH params and Alpha-Enhancement} shows that [O/Fe]$_{\mathrm{gal}}$ correlates mostly with $t_{\mathrm{peak}}$ and is only weakly dependent on the width of the star formation history. As remarked in \S~\ref{sec:Evolution}, the galactic $\alpha$-enhancement is dominated by younger stars, and hence to obtain a large [O/Fe]$_{\mathrm{gal}}$ the galaxy must form as little iron as possible such that the younger stars are not too $\alpha$-suppressed. The peak time is much more important than the width of the SFH in terms of the total synthesized iron until today, with larger peak times corresponding to less iron and higher galactic $\alpha$-enhancement. A second-order effect is that, once the peak time is already large, a small $\sigma_{\mathrm{SFR}}$ further reduces the iron buildup and enhances [O/Fe]$_{\mathrm{gal}}$. Correspondingly, to obtain a small galactic $\alpha$-enhancement, the galaxy needs to synthesize as much iron as early as possible, which is achieved via small peak times and (as second-order effect) small SFH widths. This reasoning accounts for all features in Figure~\ref{fig:SFH params and Alpha-Enhancement}. 

We also note that the right panel in Figure~\ref{fig:SFH params and Alpha-Enhancement} explains why at fixed mass, the galactic $\alpha$-enhancement correlates with sSFR (see Figure~\ref{fig:MainSequence} and \citealt{Matthee2018}). Compared to other galaxies with the same mass, galaxies with a high sSFR at $z=0$ tend to have had a relatively high sSFR for about half the age of the Universe \citep{Matthee2019}. Therefore, precisely because we look at galaxies with fixed mass, galaxies with a high sSFR have a SFH with a late peak, and consequently a high galactic $\alpha$-enhancement.

\section{Impact on Integrated Spectra}\label{sec:Impact}

Here we investigate the impact of variations in $\alpha$-enhancements on the integrated galaxy spectra of star-forming galaxies. While the $\alpha$-enhancement of passive galaxies has been measured routinely, the corresponding measurements for star-forming galaxies are relatively unexplored territory. This is because the relevant spectral absorption features are typically much weaker in galaxies with younger stellar populations compared to passive galaxies \citep[e.g.][]{Conroy2014}.

Recently, spatially resolved spectroscopy \citep[e.g.][]{Neumann2020} has allowed measurements of $\alpha$-enhancement in (parts of) relatively nearby star-forming galaxies. Moreover, \cite{Gallazzi2021} presented measurements of $\alpha$-enhancement in SDSS fibre spectra of the centers of star-forming galaxies. However, it is unclear whether the methodology (i.e. calibrations of combinations of Lick indices) that has been used for passive galaxies can be applied to star-forming galaxies. In this section we therefore explore the character and strength of the expected spectral signatures depending on [O/Fe]$_{\mathrm{mean}}$ and [O/Fe]$_{\mathrm{gal}}$.

\begin{figure*}
	\includegraphics[width=0.95\textwidth]{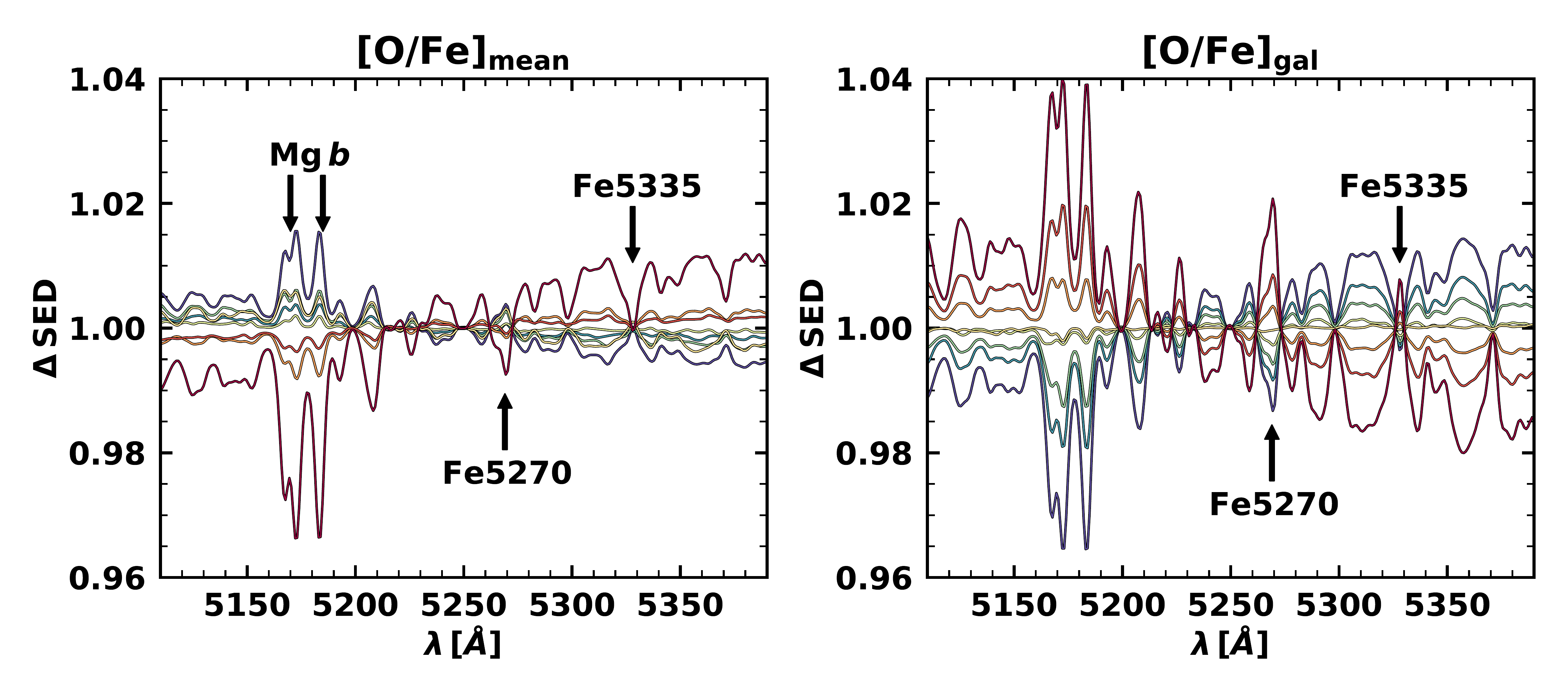}
    \caption{The impact of variations in $\alpha$-enhancements in the median spectra of star-forming galaxies. The left panel shows variations in [O/Fe]$_{\mathrm{mean}}$ and the right panel shows variations in [O/Fe]$_{\mathrm{gal}}$. Each line shows the spectrum of an $\alpha$-enhancement bin (equivalent color-coding as in Figure~\ref{fig:SFHs}, red being the most enhanced and blue the least) relative to the median spectrum of all star-forming galaxies. Shown is the wavelength region containing the Mg$\,b$ (around 5170\,\AA) and the Fe5270 and Fe5335 features.}
    \label{fig:SED_ratios}
\end{figure*}

\subsection{Constructing synthetic spectra}
As a simple approach to model SEDs of EAGLE galaxies, we use the MILES SSP models (\citealt{Vazdekis2015}) which cover the wavelength range 3525-7500\,{\AA} (binned to 0.9\,\AA) to model the SEDs of the star particles in the simulated galaxies. MILES is an empirical library containing self-consistent $\alpha$-enhanced models for the BaSTI isochrones which take the impact of [$\alpha$/Fe] on the isochrones into account. We use a Chabrier IMF (\citealt{Chabrier2003}) throughout for consistency with EAGLE (we note that MILES and EAGLE use the same stellar mass interval of $0.1-100\,\mathrm{M}_{\odot}$). We interpolate the age, metallicity and $\alpha$-enhancement of the star particles (with initial masses M$_{\rm star}=10^6$ M$_{\odot}$) in the simulated galaxies to the MILES parameter grid which spans ages from 0.03 to 14\,Gyr (53 values), metallicities from -2.27 to 0.4 (12 values) and two values of $\alpha$-enhancement (0 and +0.4). Unfortunately, there are only very few self-consistent SSP libraries with an extended range of $\alpha$-enhancements. While having only two possible $\alpha$-enhancements for the SSP models is not optimal, the global $\alpha$-enhancements of the galaxies (i.e. [O/Fe]$_{\mathrm{mean}}$ and [O/Fe]$_{\mathrm{gal}}$) are still captured since we average over $>1000$ star particles per galaxy (see also \citealt{Pinna2019}). We do not expect that absorption from interstellar dust can significantly alter our results since we focus on absorption line indices in a narrow wavelength region (see \citealt{Trayford2017} and \citealt{Trcka2020} for studies of dust affecting the SEDs of EAGLE galaxies), hence we do not model dust attenuation for our calculation of galactic SEDs. We also do not include nebular emission. Since the total luminosities of the galaxies are correlated with the $\alpha$-enhancements we normalise all SEDs by their integrated luminosity over the wavelength region of the magnesium and iron features between $(5110-5390)\,$\AA.

\subsection{Imprints on Mg$\,b$, Fe5270 and Fe5335} 
The relative strengths of the Mg$\,b$, Fe5270 and Fe5335 Lick indices are known to be sensitive to variations in [$\alpha$/Fe] \citep[e.g.][]{McQuitty1994,Trager1998}. Recently, \citet{Goncalves2020} indeed showed that these features lie in the wavelength region that is most optimal for measuring [$\alpha$/Fe] in integrated spectra. Hence, for this short discussion, we focus on the wavelength region encompassing these features.

For the analysis of how [O/Fe]$_{\mathrm{mean}}$ and [O/Fe]$_{\mathrm{gal}}$ leave imprints on the SEDs, we bin the galaxies in our sample into eight subsets ordered by $\alpha$-enhancement (as done in \S~\ref{sec:SFHs}). We calculate median SEDs for each of these subsets and divide these by the median SED of our entire galaxy sample. This therefore shows the change in the SED as one varies the $\alpha$-enhancement. The result of this calculation is shown in Figure~\ref{fig:SED_ratios}, with the same color-coding as in Figure~\ref{fig:SFHs}. It is clear that variations in $\alpha$-enhancement, despite that these variations are small, can be seen in the integrated spectra of star-forming galaxies. Comparing the two definitions of $\alpha$-enhancement, we see that the changes in the SEDs with increasing [O/Fe] behave oppositely, similar to the behaviour of the variations in their SFHs shown in Figure~\ref{fig:SFHs}. The differences are larger for variations in [O/Fe]$_{\mathrm{gal}}$ because we have seen that the galactic $\alpha$-enhancement is impacted relatively more by younger stars which dominate the SED.

We notice that the Mg$\,b$, Fe5270 and Fe5335 lines clearly vary with varying $\alpha$-enhancement, with variations of $\approx3\,\%$ between the most extreme SEDs (which correspond to variations of 0.05 and 0.09 dex in [O/Fe]$_{\mathrm{mean}}$ and [O/Fe]$_{\mathrm{gal}}$, respectively). These variations are therefore second-order effects (as the strengths of the absorption lines are on the order of 20 \%), but they are observable when the signal-to-noise ratio of spectra is $>30$ per resolution element of 2.5 {\AA} (which is the FWHM resolution of the MILES spectra that we use).

\begin{figure*}
	\includegraphics[width=0.95\textwidth]{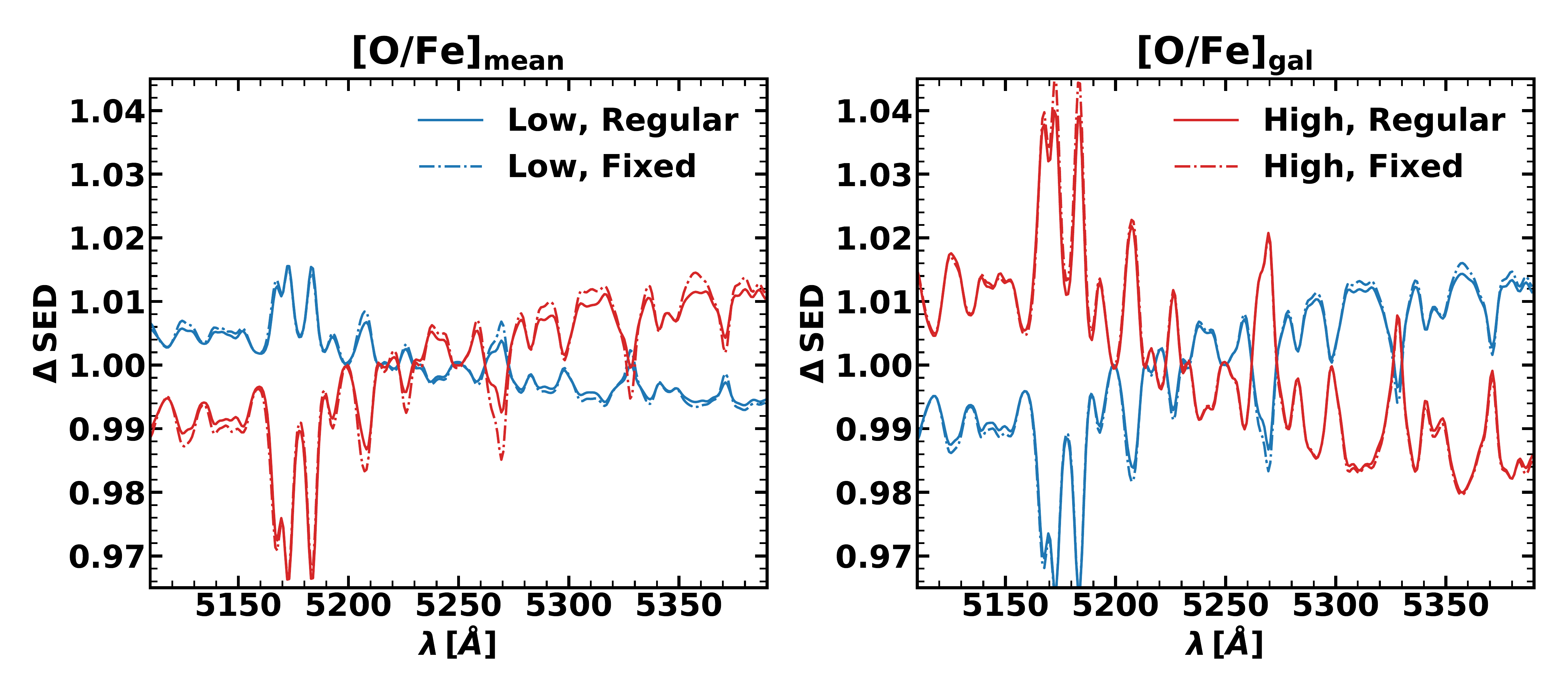}
    \caption{Median SEDs (as in Figure~$\ref{fig:SED_ratios}$) of the two subsets with the highest (red) and lowest (blue) $\alpha$-enhancements for [O/Fe]$_{\rm mean}$ (left) and [O/Fe]$_{\rm gal}$ (right), respectively. The solid lines show the synthetic SEDs taking the actual [$\alpha$/Fe] of the star particles into account, while the dash-dotted lines show the synthetic SEDs assigning all star particles a fixed solar $\alpha$-enhancement.}
    \label{fig:SEDs_CheckAlphaImpact}
\end{figure*}

To address whether the observed variations in the spectra are due to variations in the $\alpha$-enhancements of stellar atmospheres, we perform the following test. We create alternative synthetic SEDs of the simulated galaxies where we fix the [$\alpha$/Fe] of the star particles to the solar value, instead of using the actual value. Then, we repeat the analysis and show the relative changes in the SEDs when varying the mean and the galactic $\alpha$-enhancement. The results are shown in Figure~\ref{fig:SEDs_CheckAlphaImpact}. For visualisation purposes we only show the normalised SEDs of the high and low extremes of the $\alpha$-enhancement distributions. The dot-dashed (fixed [$\alpha$/Fe]) and solid (varying [$\alpha$/Fe]) lines are basically indistinguishable with differences on the $\lesssim 0.4$ \% level. This means that most of the variations in the synthetic spectra that correlate with variations in [$\alpha$/Fe] are not caused by them, but rather trace underlying correlations between the absorption lines and the stellar age and metallicity distributions.

\subsection{Impact on Mg\,$b/\langle\mathrm{Fe}\rangle$}
In integrated galaxy spectra the measurement of $\alpha$-enhancement has often been based on Mg\,$b/\langle\mathrm{Fe}\rangle$ \citep[e.g.][]{Thomas2003, Gallazzi2021}, where $\langle\mathrm{Fe}\rangle=0.5(\mathrm{Fe}5270+\mathrm{Fe}5335)$. We explore how well this index ratio (ranging from 0.50 to 0.72 for our star-forming galaxy sample) correlates with [O/Fe]$_{\mathrm{mean}}$, [O/Fe]$_{\mathrm{gal}}$ and other properties of the stellar populations in simulated galaxies. In addition to the mass-weighted quantities we have been using so far we also show the correlation coefficients for the corresponding light-weighted quantities, such that for instance the stellar masses in Eqns.~\ref{eq:OFe_mean} and~\ref{eq:OFe_gal} are replaced by stellar luminosities (integrated over the MILES wavelength range of 3525-7500\,\AA). We explore both the regular SEDs and the SEDs calculated with $\alpha$/Fe fixed to the solar value. The results are listed in Table $\ref{tab:MgbFe}$. 

\begin{table}
    \centering
    \caption{Pearson-r correlation coefficients between the Mg\,$b/\langle\mathrm{Fe}\rangle$ index and properties of the stellar populations in our sample of star-forming galaxies. We investigate both mass-weighted (MW) and light-weighted (LW) quantities and show the results for the regular synthetic SEDs where variations in the stellar [$\alpha$/Fe] are accounted for and for the synthetic SEDs where [$\alpha$/Fe] is fixed to the solar value. }
    \begin{tabular}{c|c|c|c|c|c}
        SED type&Weights&Age&Z&[O/Fe]$_{\mathrm{mean}}$&[O/Fe]$_{\mathrm{gal}}$ \\ \hline
         \multirow{2}{*}{Regular}& MW & 0.45 & 0.30 & 0.46 & -0.14\\
         & LW & 0.74 & 0.21 & 0.56 & -0.47 \\ \hline
         \multirow{2}{*}{Fix. [$\alpha$/Fe]}& MW & 0.77 & 0.71 & 0.24 & -0.63\\
         & LW & 0.92 & 0.62 & 0.14 & -0.84
    \end{tabular}
    \label{tab:MgbFe}
\end{table}

\begin{figure}
	\includegraphics[width=\columnwidth]{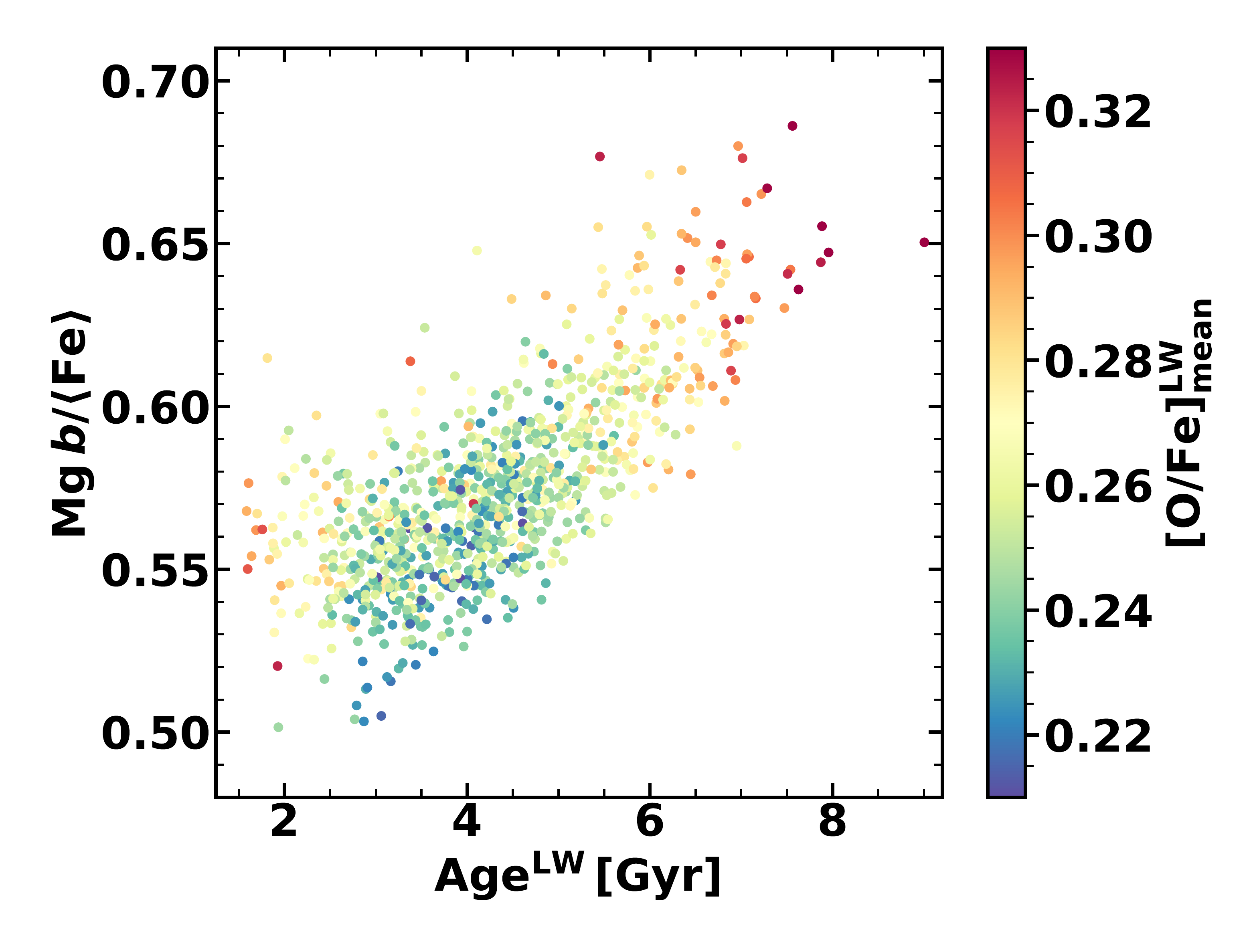}
    \caption{The correlation between the Mg\,$b/\langle\mathrm{Fe}\rangle$ index and the light-weighted age in the sample of star-forming galaxies using synthetic spectra that take variations in [$\alpha$/Fe] into account. The colour-coding shows the light-weighted [O/Fe]$_{\rm mean}$ and demonstrates that there are no significant variations in the spectral index that correlate with [O/Fe]$_{\rm mean}$ at fixed age. We limit the range of the colorbar and exclude one galaxy with peculiarly low age for optimal visualisation.}
    \label{fig:Lick}
\end{figure}

From Table $\ref{tab:MgbFe}$ it is clear that, for our sample of star-forming galaxies, Mg\,$b/\langle\mathrm{Fe}\rangle$ varies mostly with the light-weighted age of the stellar populations. The index is not strongly correlated with mass-weighted $\alpha$-enhancements, but there is a mild correlation with the light-weighted $\alpha$-enhancements. 
These correlations are however not independent of the light-weighted age. Indeed, Figure~\ref{fig:Lick} shows that there is no evidence for variations in Mg\,$b/\langle\mathrm{Fe}\rangle$ that correlate with variations in the $\alpha$-enhancement and are independent of variations in the light-weighted age. This demonstrates that the variations in Mg\,$b/\langle\mathrm{Fe}\rangle$ in our simulated spectra are mostly due to variations in age. 

We can learn more about this result by looking at the correlations between Mg\,$b/\langle\mathrm{Fe}\rangle$ and the properties in the SEDs computed with fixed [$\alpha$/Fe] (lower part in Table~\ref{tab:MgbFe}). The correlation coefficients are now even stronger with age, metallicity and the $\alpha$-enhancements. This implies that the spectral variations induced by differences in the [$\alpha$/Fe] abundances in the stellar atmospheres are somewhat in the opposite direction compared to the (much larger) spectral variations due to age and metallicity.

We stress that these results do not necessarily mean that measurements of the Mg\,$b/\langle\mathrm{Fe}\rangle$ index are always measuring age variations instead of $\alpha$/Fe variations -- it depends of the actual amount of variation in [$\alpha$/Fe]. Besides the correlation coefficients between the Mg\,$b/\langle\mathrm{Fe}\rangle$ index and the properties listed in Table $\ref{tab:MgbFe}$, we find that the average value (i.e. the normalisation) of Mg\,$b/\langle\mathrm{Fe}\rangle$ is higher by $\approx0.12$ in the regular SEDs compared to the SEDs that are constructed with solar [$\alpha$/Fe]. The difference between the solar [$\alpha$/Fe] of zero and the median [$\alpha$/Fe] in our star-forming sample are +0.3 and +0.2 dex for the mean and galactic $\alpha$-enhancement, respectively. This means that variations in [$\alpha$/Fe] that are on the order $>0.2$ dex do have a significant imprint on the value of Mg\,$b/\langle\mathrm{Fe}\rangle$ and can thus be measured \citep[e.g.][]{Gallazzi2021}. However, the variations of [$\alpha$/Fe] among the simulated star-forming galaxies in our study are much smaller ($\sigma\approx0.03$ dex). In this regime, we thus conclude that the changes in Mg\,$b/\langle\mathrm{Fe}\rangle$ are particularly due to variations in the age distribution. Therefore, this implies a systematic lower limit of the degree of variations in [$\alpha$/Fe] of $\approx0.1$ dex that can be measured in integrated spectra. Our result also implies that SED fitting codes that use a single fixed value of [$\alpha$/Fe] \citep[e.g.][]{Robotham2020} are not heavily affected by this assumption for star-forming galaxies.

\section{Discussion: comparison to passive galaxies}\label{sec:Discussion}
We have seen that while variations in $\alpha$-enhancement are correlated with variations in the observed spectra of star-forming galaxies, these mostly trace differences in the age and metallicity distributions of stars. As comparison and discussion, here we also perform several parts of our analysis for central passive galaxies with masses above $M_*=3\times10^{10}$ M$_{\odot}$ which are devoid of star formation.

\subsection{SEDs}

\begin{figure*}
	\includegraphics[width=0.95\textwidth]{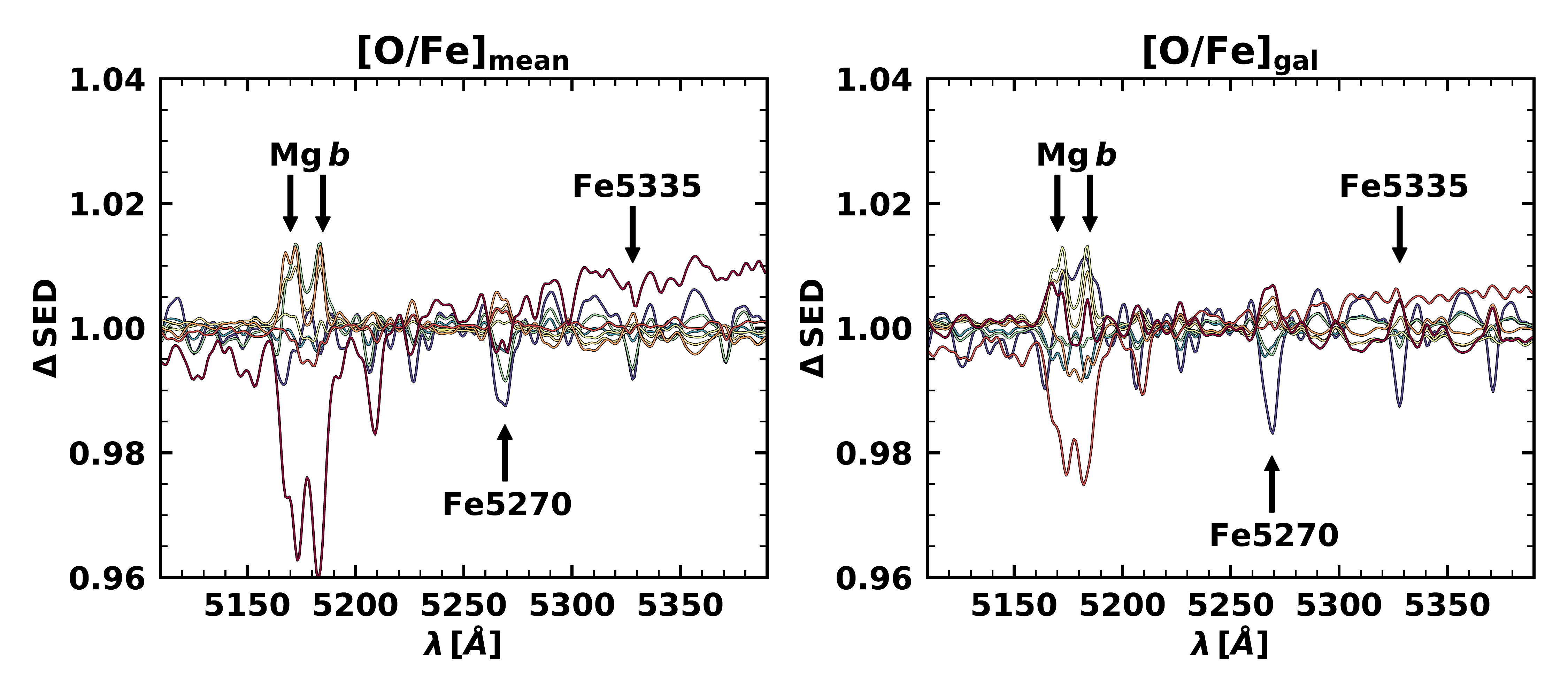}
    \caption{Same as Figure~\ref{fig:SED_ratios}, but for the passive sample (see Table~\ref{tab:GalaxySample}). The left panel shows variations in [O/Fe]$_{\mathrm{mean}}$ and the right panel shows variations in [O/Fe]$_{\mathrm{gal}}$. Each line shows the spectrum of an $\alpha$-enhancement bin (equivalent color-coding as in Figure~\ref{fig:SFHs}, red being the most enhanced and blue the least) relative to the median spectrum of all passive galaxies. Shown is the wavelength region containing the Mg$\,b$ (around 5170\,\AA) and the Fe5270 and Fe5335 features.}
    \label{fig:SED_ratios_passive}
\end{figure*}

We repeat the calculation of the synthetic spectra for subsets of galaxies split by $\alpha$-enhancement for this passive sample, and show the result in Figure~\ref{fig:SED_ratios_passive}. Although the contrast in the absorption features is still just $\approx3\,\%$, comparable to the star-forming sample, the retrieval of galactic $\alpha$-enhancements appears more feasible. Due to the positive correlation between the mean and galactic $\alpha$-enhancement for passive galaxies (see Figure~\ref{fig:AlphaPlane}), the spectral variations in both panels coincide. Unlike the synthetic spectra of star-foming galaxies, we notice that the absorption features from magnesium and iron lines are now anti-correlated in the expected way, with $\alpha$-enhanced galaxies having comparatively large magnesium and weak iron lines, respectively.

\subsection{Correlations between $\mathrm{Mg}\,b/\langle\mathrm{Fe}\rangle$ and galaxy properties}

We investigate the correlations between the Mg$\,b/\langle\mathrm{Fe}\rangle$ index and various galactic properties for the sample of passive galaxies. The index varies from 0.70 to 0.89 for the 27 galaxies in this sample, to be compared to a range from 0.50 to 0.72 for the galaxies in the star-forming sample. The Pearson-r coefficients for the correlations of $\mathrm{Mg}\,b/\langle\mathrm{Fe}\rangle$ with various galactic properties are listed in Table~\ref{tab:MgbFe_passive}. In stark contrast to the star-forming sample, the passive sample has a moderately (strong) correlation between [O/Fe]$_{\mathrm{mean}}$ ([O/Fe]$_{\mathrm{gal}}$) and Mg$\,b/\langle\mathrm{Fe}\rangle$ which vanishes when calculating the synthetic SEDs with a fixed [$\alpha$/Fe] for the stellar atmospheres. This demonstrates that the observed variations in the index are indeed due to variations in the [$\alpha$/Fe] distribution of stars and not due to confounding variables such as stellar age. It is interesting to note that the index correlates most strongly with [O/Fe]$_{\mathrm{gal}}$, which we have shown that can be interpreted as the metallicity-averaged [$\alpha$/Fe] of the galaxy. This is in line with our expectation in \S~\ref{sec:AverageAlpha} that the $\alpha$-enhancement measured via Mg$\,b/\langle\mathrm{Fe}\rangle$ corresponds to [O/Fe]$_{\mathrm{gal}}^{\mathrm{LW}}$ (Eqn.~\ref{eq:obs1}).

\begin{table}
    \centering
    \caption{Pearson-r correlation coefficients between the Mg\,$b/\langle\mathrm{Fe}\rangle$ index and properties of the stellar populations in our sample of passive galaxies (see Table~\ref{tab:GalaxySample}).}
    \begin{tabular}{c|c|c|c|c|c}
        SED type&Weights&Age&Z&[O/Fe]$_{\mathrm{mean}}$&[O/Fe]$_{\mathrm{gal}}$ \\ \hline
         \multirow{2}{*}{Regular}& MW & 0.40 & 0.52 & 0.43 & 0.75\\
         & LW & 0.50 & 0.40 & 0.51 & 0.85 \\ \hline
         \multirow{2}{*}{Fix. [$\alpha$/Fe]}& MW & 0.19 & 0.89 & -0.02 & 0.11\\
         & LW & 0.26 & 0.89 & 0.06 & 0.25
    \end{tabular}
    \label{tab:MgbFe_passive}
\end{table}

\subsection{Constraining SFHs through $\alpha$-enhancement}

For passive galaxies, a wealth of studies have described the connection between star formation histories and $\alpha$-enhancement. In a pioneering study, \citet{Thomas2005} quantitatively describe this connection as $[\alpha/\mathrm{Fe}]\approx1/5-1/6\log_{10}{\Delta t}$, where $[\alpha/\mathrm{Fe}]$ is the $V$-band luminosity-weighted mean $\alpha$-enhancement\footnote{Technically, \citet{Thomas2005} use a slightly different quantity than the one defined in Eqn.~\ref{eq:OFe_mean} as they swap the operational order of averaging and logarithm (corresponding to Eqn.~\ref{eq:obs2}, but with luminosity instead of mass weights).} (see their section 6.1) and $\Delta t$ corresponds to the FWHM of a SFH characterised by a Gaussian. Figure~\ref{fig:SFH params and Alpha-Enhancement} displays a very similar relation, where the $\alpha$-enhancements of galaxies are plotted in the $t_{\mathrm{peak}}$-$\sigma_{\mathrm{SFR}}$ plane. Here we fit the following linear relation of SFH parameters to the light-weighted [O/Fe]$_{\mathrm{mean}}^{\mathrm{LW}}$ and [O/Fe]$_{\mathrm{gal}}^{\mathrm{LW}}$:

\begin{equation}
    \mathrm{[O/Fe]_{mean/gal}^{LW}}=a\log_{10}t_{\mathrm{peak}}+b\log_{10}\sigma_{\mathrm{SFR}}+c,
    \label{eq:PredictAlpha}
\end{equation}

where the SFH parameters are measured in Gyr. We use the \texttt{python} \texttt{lmfit} package to find the free parameters $a$, $b$ and $c$ and their uncertainties in Eqn.~\ref{eq:PredictAlpha}. The resulting parameters for both the star-forming and the passive sample are listed in Table~\ref{tab:PredictAlpha}. The results for the relation between SFH parameters and $\alpha$-enhancements are consistent with those presented in Figures~\ref{fig:SFHs} and~\ref{fig:SFH params and Alpha-Enhancement} for the star-forming sample. We remark that the linear approximation of Eqn.~\ref{eq:PredictAlpha} is a poor description for the mean $\alpha$-enhancement of the star-forming sample. 

For the light-weighted mean $\alpha$-enhancement of the passive sample, which is the quantity closest to the one used by \citet{Thomas2005}, we find a $\sigma_{\mathrm{SFR}}$\footnote{We remark that $\sigma_{\mathrm{SFR}}$ is defined as the FWHM of a log-normal SFH, and hence comparable to the FWHM $\Delta t$ of a Gaussian SFH used by \citet{Thomas2005}.} slope of $b=-0.20$ which is very comparable to $-1/6$. We therefore confirm the canonical result that variations in $[\alpha$/Fe] correlate with the compactness of the star formation history. Additionally, we find a $t_{\mathrm{peak}}$ slope of $a=-0.19$, indicating that early-peaked SFHs additionally lead to large mean $\alpha$-enhancements for passive galaxies. 

For [O/Fe]$_{\mathrm{gal}}^{\mathrm{LW}}$, we find a strong negative correlation with $\sigma_{\mathrm{SFR}}$ ($b=-0.38$), while the correlation with $t_{\mathrm{peak}}$ is not significant. As we have seen that the Mg\,$b/\langle\mathrm{Fe}\rangle$ index most strongly correlates with [O/Fe]$_{\mathrm{gal}}^{\mathrm{LW}}$, this result therefore confirms that the {\it observed} spectral variations in passive galaxies are mostly sensitive to variations in [$\alpha$/Fe] that are caused by different widths of the star formation history and not necessarily the peak time.

\begin{table}
    \centering
    \caption{Best-fitting parameters to predict light-weighted mean and galactic $\alpha$-enhancement from SFH parameters (Eqn.~\ref{eq:PredictAlpha}) for the samples of star-forming galaxies and passive galaxies. $a$ and $b$ denote the correlations with $t_{\mathrm{peak}}$ and $\sigma_{\mathrm{SFR}}$, $c$ is the normalisation.}
    \begin{tabular}{c|c|c|c}
        Quantity&$a$&$b$&$c$\\ \hline
        SFGs & & & \\
          {[O/Fe]}$_{\mathrm{mean}}^{\mathrm{LW}}$ & $0.025\pm0.007$ & $-0.036\pm0.005$ & $0.274\pm0.004$\\
          {[O/Fe]}$_{\mathrm{gal}}^{\mathrm{LW}}$ & $0.094\pm0.010$ & $0.095\pm0.006$ & $-0.038\pm0.005$ \\ \hline
        Passive & & & \\

          {[O/Fe]}$_{\mathrm{mean}}^{\mathrm{LW}}$ & $-0.19\pm0.05$ & $-0.20\pm0.04$ & $0.61\pm0.02$ \\
          {[O/Fe]}$_{\mathrm{gal}}^{\mathrm{LW}}$ & $0.11\pm0.09$ & $-0.38\pm0.07$ & $0.39\pm0.04$ \\
    \end{tabular}
    \label{tab:PredictAlpha}
\end{table}

\pagebreak

\section{Future directions}\label{sec:Future}
Here we discuss the limitations of our paper and further directions in which our understanding of variations in $\alpha$-enhancements can be improved. 

On the simulation side, there are several limitations to our study. First, our results rely on a single hydrodynamical simulation with a sub-grid model for stellar yields and the delay time distribution of Type Ia supernovae \citep{Wiersma2009b}. Therefore our results could quantitatively change somewhat in case any of these models can be improved. To first order, we expect small effects. Different yields would only change the normalisation of [$\alpha$/Fe] (\citealt{Larson1972}; \citealt{Weinberg2017}) and a different delay time distribution may lead to a slightly different relation between age, $Z$ and [$\alpha$/Fe] (\citealt{Weinberg2017}). Second, the IMF is assumed to be fixed. In case the IMF would vary, the relative amounts of Type II to Type Ia SNe could change, impacting the $\alpha$-enhancement (e.g. \citealt{DeMasi2018}; \citealt{Barber2019}). Such variations could be quite complicated, as IMF variations may be (indirectly) correlated with the SFH. This could be addressed in a future study. Nevertheless, we stress that the EAGLE simulation matches the observed evolution of the cosmic SN Ia density \citep{Schaye2015}.

On the analysis side, we note that we have only parametrised the star formation history, and not the chemical evolution history. One would expect that the chemical evolution is tied to the star formation history (see for instance a model that is used in the ProSpect SED generator code; \citealt{Robotham2020}). However, we note that such a connection may not be straightforward because chemical evolution history is also influenced by metal-enriched gas in- and outflows. These on themselves could be correlated with the SFH in a complex way. A parametrised connection between the SFH and chemical enrichment is appealing particularly as it could be used as prior for fitting of observed spectra. 

In this paper we only focused on the $\alpha$-enhancements of the full simulated galaxy (i.e. subhalo). It could be interesting to study spatial gradients in $\alpha$-enhancement as they will yield information on the way galaxies build-up. In particular it will be interesting to focus on the bulges of star-forming galaxies, as it will likely be easier to observe $\alpha$-variations there than in the spiral arms (e.g. \citealt{Neumann2020}). 

Finally, in the future the synthetic SEDs can be improved in various ways. Improved libraries with a finer grid in [$\alpha$/Fe] ratios would be welcomed (as in the recent sMILES library, \citealt{Knowles2021}). It is possible that short-term burstiness of the SFHs is underestimated due to the resolution of  the EAGLE simulation \citep[e.g.][]{Iyer2020}. This could result in an underestimate of the contribution of the flat-continuum spectra from youngest stars to the integrated spectra and therefore result in overestimated absorption line strengths. Furthermore the effects of relative dust attenuation of young and old stars could be investigated (e.g. \citealt{Camps2016}, \citealt{Trcka2020}).

\section{Summary}\label{sec:Conclusions}

In this paper we investigated the origin of variations in stellar $\alpha$-enhancements of star-forming galaxies in the largest reference simulation of the EAGLE project in the present-day Universe. Oxygen is used as a proxy for the $\alpha$-abundance throughout our paper. We investigated two distinct definitions of $\alpha$-enhancement, their connection to the star formation and chemical enrichment histories and their impact on integrated synthetic spectra. We discussed our results in the context of interpreting observed spectra and compared our results to the previously studied $\alpha$-enhancement variations in passive galaxies. The following points summarize our main findings:

\begin{itemize}
    \item The stellar $\alpha$-enhancement of a composite stellar population is not uniquely defined. We present two definitions of $\alpha$-enhancement: mean $\alpha$-enhancement [O/Fe]$_{\mathrm{mean}}$ (the mass-weighted average oxygen-to-iron ratio, Eqn.~\ref{eq:OFe_mean}) and galactic $\alpha$-enhancement [O/Fe]$_{\mathrm{gal}}$ (the ratio of the galactic oxygen abundance to the galactic iron abundance, Eqn.~\ref{eq:OFe_gal}). Both definitions of $\alpha$-enhancement of galaxies correspond to the same quantity, the stellar oxygen-to-iron ratio, but they are weighted differently. For [O/Fe]$_{\mathrm{mean}}$, each star (or SSP) is weighted by its current mass, while for [O/Fe]$_{\mathrm{gal}}$, each star is weighted by its current \textit{iron} mass.
    
    \item Applying these two definitions to a mass-limited sample ($5\times10^9\,\mathrm{M_{\odot}}<M_*<10^{10}\,\mathrm{M_{\odot}}$) of central star-forming galaxies from the EAGLE simulation, we find that [O/Fe]$_{\mathrm{mean}}$ and [O/Fe]$_{\mathrm{gal}}$ are slightly anti-correlated (Figure~\ref{fig:AlphaPlane}). For a high-mass sample ($M_*>3\times10^{10}\,\mathrm{M_{\odot}}$) of passive central galaxies, the two definitions correlate strongly. 
    
    \item We find, in accordance with previous studies targeting the $\alpha$-enhancement of passive galaxies, that star formation histories are crucial in determining the $\alpha$-enhancements of the star-forming galaxies. When splitting the star-forming galaxies according to their [O/Fe]$_{\mathrm{mean}}$ or [O/Fe]$_{\mathrm{gal}}$, we discover a distinct trend such that a galaxy with a high-[O/Fe]$_{\mathrm{mean}}$ or a low-[O/Fe]$_{\mathrm{gal}}$ has an early and rapid star formation history and vice versa (Figure~\ref{fig:SFHs}).
    
    \item The reversed dependence of the two definitions of $\alpha$-enhancement on the SFH can be explained when taking both the galactic chemical evolution (Figure~\ref{fig:CEHs}) and the different weighting functions (Figure~\ref{fig:Weighting Functions}) into account. [O/Fe]$_{\mathrm{mean}}$ follows the well-known trend of passive galaxies such that early-peaked, compact SFHs lead to a high $\alpha$-enhancement. However, the iron-weighting for [O/Fe]$_{\mathrm{gal}}$ leads to young stars dominating this quantity. Hence, to acquire a large galactic $\alpha$-enhancement, it is more `efficient' to delay star formation in order to boost the oxygen-to-iron ratio of the youngest stars instead of having a very rapid SFH.
    
    \item By fitting individual star formation histories with log-normal functions, we explore the correlation of the $\alpha$-enhancements of star-forming galaxies with SFHs at the level of individual galaxies. Interestingly, [O/Fe]$_{\mathrm{mean}}$ is mostly correlated with the width of the star formation history $\sigma_{\mathrm{SFR}}$, while [O/Fe]$_{\mathrm{gal}}$ is mostly correlated with the peak time $t_{\mathrm{peak}}$ (Figure~\ref{fig:SFH params and Alpha-Enhancement}). For [O/Fe]$_{\mathrm{mean}}$, the correlation fades at the low-[O/Fe]$_{\mathrm{mean}}$-end.
    
    \item We investigate the impact of [$\alpha$/Fe] variations on synthetic integrated spectra by mapping the star particles in simulated galaxies to SSP models with different levels of $\alpha$-enhancement and metallicity from MILES (\citealt{Vazdekis2015}). We show that variations in well-known absorption lines as Mg$b$, Fe5270 and Fe5335 correlate with variations in $\alpha$-enhancements (Figure~\ref{fig:SED_ratios}). However, the spectral differences can be attributed to variations in the age distribution of the star particles, and they are not due to [$\alpha$/Fe] variations of the stellar atmospheres themselves (Figure~\ref{fig:Lick}). This is because, for the dynamic range of [$\alpha$/Fe] variations in our simulated star-forming galaxies, spectra are not significantly affected. If star-forming galaxies have larger, $>0.1$ dex differences in their $\alpha$-enhancements they may have an observable imprint on Mg\,$b/\langle\mathrm{Fe}\rangle$ that is independent of the stellar ages and metallicities. 
    
    \item Finally, as a comparison, we perform a similar analysis for simulated passive galaxies (\S $\ref{sec:Discussion}$). We confirm earlier results that variations in $\alpha$-enhancements in these galaxies correlate with the compactness of the star formation history and that these variations cause spectral variations of, for example, the Mg\,$b/\langle\mathrm{Fe}\rangle$ index.

\end{itemize}

This work shows that the star formation histories of star-forming galaxies determine their $\alpha$-enhancements, similar as passive galaxies. However, results for the $\alpha$-enhancement of passive galaxies cannot be extended to star-forming galaxies in a straightforward fashion. We stress that both for observational and theoretical studies of $\alpha$-enhancement in the context of star-forming galaxies, [$\alpha$/Fe] should be carefully defined. 

With respect to resolved measurements of age, metallicity and $\alpha$-enhancement simultaneously in stars in very local galaxies, our results imply that both the width and the peak time of the SFH can be retrieved by weighting the individual $\alpha$-enhancements differently. Ultimately variations in SFHs can be related to variations in dark matter accretion histories on long timescales, and therefore future measurements of the variation of $\alpha$-enhancements among star-forming galaxies can provide unique information on the growth of structure on the longest timescales.

\section*{Acknowledgements}
We thank our anonymous referee for the constructive feedback. We extend our gratitude to Maarten Baes, Simon Lilly, Rafael Ottersberg, Gabriele Pezzulli, Alvio Renzini and Andrea Weibel for insightful discussions. AG gratefully acknowledges financial support from the Fund for Scientific Research Flanders (FWO-Vlaanderen, project G.0G04.16N). This work used the DiRAC Data Centric system at Durham University, operated by the ICC on behalf of the STFC DiRAC HPC Facility (www.dirac.ac.uk). This equipment was funded by BIS National E-infrastructure capital grant ST/K00042X/1, STFC capital grant ST/H008519/1, and STFC DiRAC Operations grant ST/K003267/1 and Durham University. DiRAC is part of the National E-Infrastructure.

We have benefited from the data analysis tool \texttt{Topcat} (\citealt{Taylor2013}) and the programming language \texttt{Python}, including the \texttt{numpy} (\citealt{vanderWalt2011}), \texttt{matplotlib} (\citealt{Hunter2007}) and \texttt{scipy} (\citealt{Virtanen2020}) packages.

\section*{Data Availability}

All simulation data used in this work is accessible through the website of the EAGLE project (http://icc.dur.ac.uk/Eagle/). The galaxy catalogues are described by \citet{McAlpine2016} and the particle data by \citet{EAGLE2017}. The MILES SSP models spectra are publicly available (http://research.iac.es/proyecto/miles/pages/ssp-models.php).


\bibliography{main}{}
\bibliographystyle{aasjournal}



\end{document}